\documentclass[
nofootinbib,
 amsmath,amssymb,
 aps,prd
]{revtex4-2}

\usepackage[utf8]{inputenc}
\usepackage{graphicx}
\usepackage{dcolumn}
\usepackage{xcolor}
\usepackage{slashed} 
\usepackage{multirow} 
\usepackage{bbold} 
\usepackage{amsmath} 
\usepackage{bm}

\begin{document}


\title{Quark-photon vertex in confining models}

\author{C. Mena}
 \email{cstiven.mc2@gmail.com}
\author{L. F. Palhares}%
 \email{leticia.palhares@uerj.br}
\affiliation{%
Departamento de F\'isica Te\'orica,\\
Universidade do Estado do Rio de Janeiro, Rua S\~ao Francisco Xavier 524, 20550-013 Maracan\~a,\\ Rio de Janeiro, Brasil
}


\begin{abstract}

We compute the one-loop quark-photon vertex and the quark magnetic moment in three different models with an infrared-modified gluon propagator, namely: the massive (Curci-Ferrari-like) model, the Gribov-Zwanziger model, and the Refined Gribov-Zwanziger model. We show results for the $F_2$ form factor and analyze the role played by the  mass parameters associated with the confined gluon in these models. Using the framework of the Constituent Quark Model, we further construct the observable proton magnetic moment including the effects of QCD interactions generated by the different confining models. Our results show that there is no qualitative observable difference stemming from the presence of complex-conjugated poles. On the other hand, quantitative differences between the various confining models can be sizable, so that observable constraints would in principle be possible, provided one has  sufficient information about the constituent quark mass and the QCD running coupling.

\end{abstract}
%
\maketitle

\section{\label{sec:level1}Introduction}
Quantum Chromodynamics (QCD) is well-established as the theory of Strong Interactions, describing several observable phenomena at high energy experiments \cite{ParticleDataGroup:2022pth} using quarks and gluons as fundamental degrees of freedom. Despite the outstanding success of lattice QCD in computing e.g. hadron masses and other low energy properties \cite{Csikor:2003ng,BMW:2008jgk,Fodor:2012gf,BMW:2014pzb,Borsanyi:2020mff}, the infrared (IR) regime of non-Abelian gauge theories still encompasses several challenges, such as the mechanism behind quark and gluon confinement (for an overview on different approaches cf. \cite{Greensite:2011zz} and references therein).

In the IR region of non-Abelian gauge theories like QCD, na\"ive perturbation theory becomes ill-behaved due to the large running coupling and nonperturbative frameworks are called for. Besides lattice QCD, a variety of analytic and semi-analytic methods have been developed in the last decades. Among different approaches \cite{Kugo:1979gm,Roberts:1994dr,Alkofer:2000wg,Aguilar:2008xm,Fischer:2008uz,Boucaud:2008ky,Tissier:2010ts,Cyrol:2014kca,Siringo:2015gia,Frasca:2015yva,Comitini:2017zfp,Chaichian:2018cyv,Maas:2008ri}, it is worthwhile noting that (i) most studies are dedicated to analyzing quark and gluon (as well as ghost) correlation functions rather than observables and (ii) it is not uncommon to find some type of dynamically generated gluon mass scale, compatible with the deep IR gluon propagator observed on gauge-fixed Lattice data \cite{Cucchieri:2004mf,Cucchieri:2007md,Sternbeck:2007ug,Bogolubsky:2007ud,Cucchieri:2008fc,Oliveira:2012eh,Bicudo:2015rma}.

Here, we concentrate on confining models that display a modified gluon propagator accounting for mass generation and our ultimate goal would be to make predictions for observables. A theoretical foundation of these models may be constructed from first-principle formulations of the functional integral of gauge fields that deal with the so-called Gribov problem \cite{Gribov:1977wm,Singer:1978dk}. Back in 1977, Gribov demonstrated the existence of multiple gauge configurations associated with the same physical fields even after gauge fixing, rendering the standard Faddeev-Popov quantization ill-defined in the nonperturbative regime. In the region where couplings and fields are small, these so-called gauge copies 
do not affect the functional integral, as corroborated by the success of the Faddeev-Popov formulation of perturbative QCD in describing high-energy scatterings. Nevertheless, in the IR, the gauge coupling and/or fields may assume large values and addressing the Gribov Problem becomes therefore important for continuum approaches.

The Gribov-Zwanziger (GZ) theory \cite{Zwanziger:1989mf,Zwanziger:1992qr,Capri:2012wx} was proposed as a partial solution, introducing a local and renormalizable theory that restricts the gauge functional integral via a gap equation (for a systematic review cf. \cite{Vandersickel:2012tz}). There are strong indications, however, that GZ theory is unstable against the formation of dimension 2 condensates \cite{Dudal:2011gd,Cucchieri:2011ig,Dudal:2019ing}, giving rise to the Refined Gribov-Zwanziger (RGZ) theory. 

Unlike the GZ theory, the RGZ tree-level gluon propagator is compatible with the deep IR behavior observed on Landau-gauge Lattice QCD data \cite{Dudal:2008sp}. Moreover, RGZ theory is a self-consistent nonperturbative formulation that can in principle be applied at all energy ranges, reducing to QCD (or pure gauge Yang-Mills, in the absence of quarks) in the ultraviolet region \cite{Capri:2015mna}. 
More recently, it has been shown \cite{Capri:2015ixa,Capri:2015nzw,Capri:2016aqq,Capri:2016gut} that RGZ displays a nilpotent Becchi-Rouet-Stora--Tyutin symmetry \cite{Becchi:1974xu,Becchi:1975nq,Tyutin:1975qk}, modified with respect to Yang-Mills theory but nevertheless allowing for the proof of gauge-parameter independence of physical predictions. In this sense, the mass scale emerging in GZ and RGZ models is totally different from the mass of a Proca field, that breaks gauge invariance and BRST symmetry in a straightforward way.

Even though both GZ and RGZ actions are constructed using fundamental gluon fields, there are no gluon asymptotic states due to the nontrivial analytical structure of the propagator in the form of complex-conjugated poles. While this could be an indication of built-in confinement, the nonstandard pole structure raises many issues when trying to define Minkowski-space quantities. The assumption is that in physical observables the standard analytical structure would be recovered. Most of the RGZ results so far involve, however, nonphysical correlation functions \cite{Capri:2015nzw,Mintz:2017qri,Mintz:2018hhx,Capri:2018ijg}, with few exceptions \cite{Dudal:2010cd,Capri:2012hh,Dudal:2013vha,Capri:2014bsa,Guimaraes:2015vra,Canfora:2013zna,Canfora:2015yia, Sumit:2023hjj,Bandyopadhyay:2023yjp,Debnath:2023dhs}.

In this paper, the aim is to test the predictions of GZ and RGZ models for a given observable, namely the anomalous magnetic moment of nucleons.
This physical quantity is particularly interesting because it is an observable sensitive to the deep IR behavior of the theory. Furthermore, high-precision measurements \cite{ParticleDataGroup:2022pth} are  available, so that developing the current analysis could lead to constraints on confining model parameters. 
Also, Lattice QCD computations of the quark-photon vertex are possible (cf. e.g. \cite{Leutnant:2018dry}) and a direct comparison of the full momentum-dependent form factor may be available in the near future.

For that matter, we compute
the quark-photon vertex, focusing on the $F_2$ form factor and the anomalous magnetic moment of quarks. To obtain the corresponding nucleon observable, we resort to the simple yet widely-used Constituent Quark Model (CQM) \cite{Perkins:1982xb, Gasiorowicz:1981jz} that should provide a good framework for a first scrutiny of confining models. For comparison, we also show results from another IR confining description: the Curci-Ferrari model \cite{Tissier:2010ts,Tissier:2011ey,Serreau:2012cg,Pelaez:2021tpq,Barrios:2020ubx,Barrios:2022hzr}.

This paper is organized as follows. In Section \ref{sec:quark-photon} we compute the quark-photon vertex and the associated $F_2$ function in Euclidean space in the general case. Then, we discuss the usual contributions that come from Quantum Electrodynamics (QED) and perturbative QCD. In section \ref{sec:results-confining}, we show full results for $F_2$ obtained in the Curci-Ferrari, GZ and RGZ confining models and discuss some interesting limits. In Section \ref{sec:nucleonMM} we briefly review the Constituent Quark Model in the description of the nucleon magnetic moment. In Section \ref{sec:confining-MM} we discuss the corrections to the proton magnetic moment brought about by the Curci-Ferrari, GZ and RGZ confining models. Finally, we show our summary and outlook in Section \ref{sec:summary}.

\section{Quark-photon vertex}
\label{sec:quark-photon}

\begin{figure}[b]
  \centering
  \begin{tabular}[b]{c}
 \fbox{ \includegraphics[width=.175\linewidth]{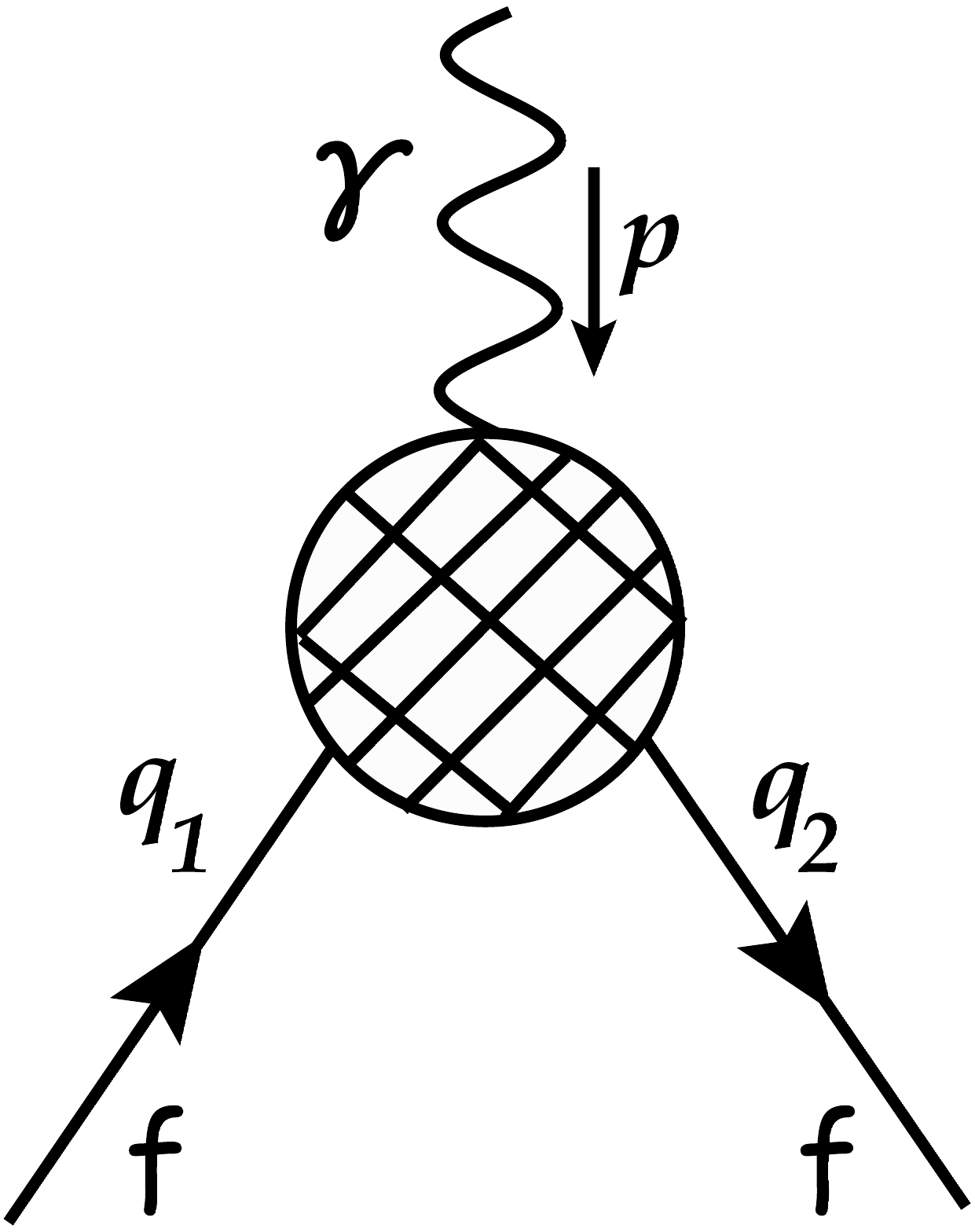}} \\
    \small (a)
  \end{tabular} \qquad
  \begin{tabular}[b]{c}
 \fbox{  \includegraphics[width=.68\linewidth]{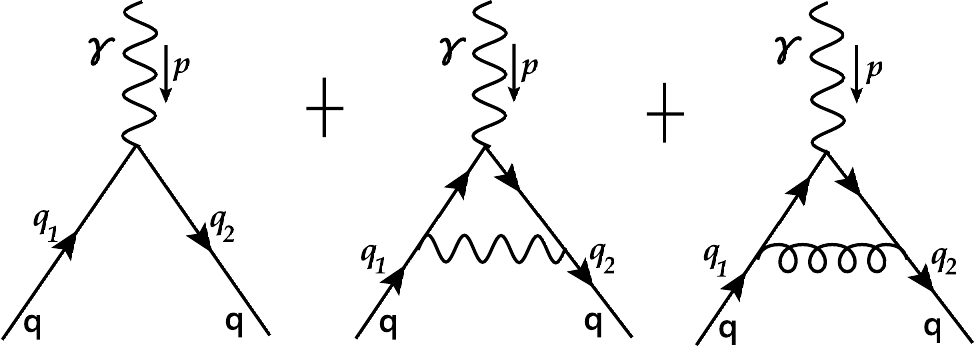}} \\
    \small (b)
  \end{tabular}
  \caption{(a) Fermion–photon vertex. (b) Quark--photon vertex up to one loop. Straight, wavy and coiled lines represent  quark, photon and gluon, respectively.}
 \label{Fig:Q_ph_Vertex_Gral}
\end{figure}

The fermion-photon vertex (Fig.  \ref{Fig:Q_ph_Vertex_Gral}.a) can be written as (cf. e.g. \cite{Peskin:1995ev,Schwartz:2014sze}):
\begin{align}
\label{Eq:AMM_GRel}
  i\mathcal{M}_{\mu} \,=-\,i\,e\, \,\overline{\mathcal{U}} (q_{2})\,  \left[ F_1\left(p^2 \right) \gamma^{\mu} +  i\,\frac{p_{\nu}\,\sigma^{\mu \nu}}{2m}\, F_2\left(p^2\right) \right] \mathcal{U} (q_{1}) ,
\end{align}
\noindent
where $\mathcal{U}(q_1)$ and $\overline{\mathcal{U}}(q_2)$ are Dirac spinors, $\sigma^{\mu \nu} \equiv \frac{1}{2}\left[\gamma^\mu,\gamma^\nu \right]$, with  $\gamma^{\mu}$ being Dirac matrices, $p_\nu=q_{2 \nu}-q_{1 \nu}$, and $m$ is the fermion mass. The functions $F_1$ and $F_2$ are standard form factors: $F_1$ is related to the renormalization of the electric charge whereas $F_2$ contributes to the magnetic moment of the fermion.

In the quark-photon vertex up to one-loop order, the contributions to the quark magnetic moment can be described by the Feynman diagrams in Fig. \ref{Fig:Q_ph_Vertex_Gral}.b.  Corrections from QED and QCD interactions are displayed by the diagrams with an internal photon propagator (wavy line) and an internal gluon propagator (coiled line), respectively. Here, we  focus on the QCD contribution to $F_2$ and analyze predictions from different confining models. The goal, as already discussed in the Introduction, is to test these models and constrain their parameters using observable information that connects directly to this process.

In Euclidean space the one-loop contributions in Fig. \ref{Fig:Q_ph_Vertex_Gral}.b can be written, in the linear covariant gauge, as\footnote{We omit the term $\sim (1-\xi)$, where $\xi$ is the linear covariant gauge parameter, since it only affects $F_1$, which is not of our current interest.}

\begin{align}
\label{Eq:AMM_Euc_1}
\mathbf{M}_\mu \,\to\,e\,Q_{q}\,\mathcal{N}\,\, \overline{U}(q_{2})
    \int \frac{d^{4} k}{(2\pi)^{4}}
    \left( D\left((k-q_1)^ 2\right)
    \frac{   \,\gamma_{\nu} ( \slashed{p} + \slashed{k} + im_q) \gamma_{\mu} ( \slashed{k} + im_q) \gamma_{\nu} \,\,
     }
    {
    [(p + k)^{2} + m_q^{2} ]\, [k^{2} + m_q^{2}]} \right) U(q_{1}) \,,
\end{align}
\noindent
where $\mathcal{N}=e^{2}\, Q_{q}^{2}$ for the diagram with the internal photon propagator and 
$\mathcal{N}=g^2\, C_{F}$ for the one with an internal gluon propagator, $eQ_q$ is the electric charge of quark $q$, $C_F$ is the color factor, and $m_q$ is the quark mass. The scalar function in the propagator of the exchanged gauge boson is represented by $D(p^2)$, being $1/p^2$ for both the photon and the perturbative, massless gluon. In the confining models we analyze here, this function will be modified by the presence of nonperturbative mass scales, becoming a sum of at most two terms of the form $\sim 1/[p^2+m_g^2]$, as will be clear in the next sections.

Following the standard procedure to compute the one-loop amplitude $\mathbf{M}_\mu$ \cite{Schwartz:2014sze} for a gauge boson propagator function of the form $D(p^2)=R/[p^2+r^2]$, one introduces Feynman parameters $x, y, z$, solves the Dirac matrix algebra, and evaluates the momentum integrals, so that one obtains the contribution proportional to $\,p_{\nu}\,\sigma_{\mu \nu}$: 
\begin{align}
\label{Eu_F2_Int_Gral}
    F_2 (p^2) \,=\, R\, \frac{\mathcal{N}}{4\pi^2} \left[ \int_{0}^{1}
dx\,dy\,dz\, \delta(x+y+z-1)\,\,
\frac{\, z(1-z)\,m_q^2}{   \Delta_E } \right],
\end{align}
where $\Delta_E\,=\, xy \,p^2 + (1-z)^2\, m_{q}^2 + z\, r^{2}$.
It is important to note that, in the GZ and RGZ computations that we present below, the $r^2$ and $R$ mass parameters will be complex in general. Working directly in Euclidean space, all standard steps in computing the triangle Feynman integral are generalized to complex mass parameters as long as the real part of $\Delta_E$ stays positive, which will always be the case. Also, since these Gribov-type confining models display a pair of complex-conjugated poles, imaginary parts are trivially cancels and the final result is real, as physically expected.

\subsection{QED contribution to $F_2$}
In QED the photon is massless, so that we have $R=1$, $\mathcal{N}=e^2\,Q_q^2$ and
\begin{align}
\Delta_{E}\,=\, xy \,p^2 + (1-z)^2\, m_{q}^2 \,.
\end{align}
Then, Eq. (\ref{Eu_F2_Int_Gral}) becomes
\begin{align}
\label{Eu_F2_Int_QED}
    F_2 (p^2) \,=\, Q_q^{2} \left(\frac{\alpha}{\pi}\right) \left[ \int_{0}^{1}
dx\,dy\,dz\, \delta(x+y+z-1)\,\,
\frac{\, z(1-z)\,m_q^2}{  xy\,p^2 + (1-z)^2 m_q^2} \right],
\end{align}
\noindent
where $\alpha=e^2/4\pi$. At $p^2=0$, the first QED correction to the quark anomalous magnetic moment becomes: 
%
\begin{eqnarray}
\label{Int_F2_zero_QEDc}
  F_2 (0)\,=\,Q_q^{2} \left(\frac{\alpha}{\pi}\right) \int_{0}^{1} \,dz \int_{0}^{1-z} \,dy\, \frac{z}{1-z} = Q_q^{2} \left(\frac{\alpha}{2\pi}\right) \,,
\end{eqnarray}
\noindent
Using $Q_e=-1$ for the electron, we recover the well-known Schwinger's result: $F_2 (0)=\alpha/2\pi$.

\subsection{Perturbative QCD contribution to $F_2$}

For the perturbative, massless gluon, we have $R=1$,  $\mathcal{N}=g^2\,C_F$, with $C_F= (N^2-1)/(2N)=4/3$, and 
\begin{align}
  \Delta_{E}\,=\, xy \,p^2 + (1-z)^2\, m_{q}^2\,,  
\end{align}
implying the same structure with a different pre-factor:
\begin{align}
\label{Eu_F2_Int_q_ph_V}
    F_2 (p^2) \,=\,C_{F}\left(\frac{\alpha_{s}}{\pi}\right) \int_{0}^{1}
dx\,dy\,dz\, \delta(x+y+z-1)\,\,
\frac{\, z(1-z)\,m_{q}^2}{  xy\,p^2 + (1-z)^2 m_{q}^2} \,,
\end{align}
\noindent
where $\alpha_s=g^2/4\pi$. At $p^2=0$:
%
%
\begin{eqnarray}
\label{Int_F2_zero_p_QCDc}
  F_2 (0)\,=\,C_F \left(\frac{\alpha_s}{2\pi}\right),
\end{eqnarray}
\noindent
which is similar to the QED contribution with $\alpha \to \alpha_s C_F$. 

Unlike in the case of QED, the QCD coupling at zero momentum is not simple to define or compute. However, many studies show it could have a finite value at zero momentum (see Ref. \cite{Deur:2016tte} and references therein). 

For the analysis of confining models that will follow, it is convenient to define a normalized $F_2$ for QCD in general as
\begin{eqnarray}
\label{Eq:Int_F2_zero_Norm}
\overline{F}_2(0)= \frac{\pi F_2(0)}{C_F\,\alpha_s} \,,
\end{eqnarray}
\noindent
such that, for the perturbative case, one has $\overline{F}_2^{\rm pert}(0)=1/2$. Confining models will naturally bring deviations from this value.

\section{Results for $F_2$ in confining models}
\label{sec:results-confining}

In this section we include the effects from confinement in the one-loop quark-photon vertex as described within the massive, GZ and RGZ models. They will modify the propagator and, therefore, yield a different form for the QCD contribution to $F_2$. The behavior brought about by each of these models can be compared as a function of ratios of their mass parameters to the quark mass.

\subsection{Massive model}
In the massive model, $\Delta_E$ is modified due to the presence of the massive term, $m_g^2$ in the gluon propagator
\begin{eqnarray}
D^{\rm Mass}\left((k-q_1)^2\right) = \frac{1}{(k-q_1)^2 + m_g^2} \,,
\end{eqnarray}
\noindent
which, in Eq.(\ref{Eu_F2_Int_Gral}) yields  $R=1$ and $\Delta_E\,=\, xy \,p^2 + (1-z)^2\, m_{q}^2 + z\, m_{g}^{2}$. The contribution to $\overline{F}_2$ in the massive model then reads:
%
\begin{align}
\label{F2_Int_pzero_V_qmgq}
\overline{F}_2^{\rm Mass}(0,a)\equiv
\frac{  \pi  F_2^{\rm Mass} (0,a) }{C_{F}\, \alpha_{s}}
\,=  \int_{0}^{1} \,dz \frac{\, z(1-z)^2}{ (1-z)^2\,  + z\,a} 
\,,
\end{align}   
\noindent
where $\overline{F}_2^{\rm Mass}(0,a)$ depends only on $a=(m^2_{g}/m^2_{q})$, i.e., the ratio between the gluon mass and the quark mass. Solving the integral above, we obtain:
\begin{align}
     \label{Eq:F2_Bar_0_Mass_a}
     \overline{F}_2^{\rm Mass}(0,a) =  \frac{1}{2} - \frac{a}{2}\left[2-(a-2) \mathrm{Log}\left[a\right]  + \frac{(a-2)^2-2}{\sqrt{a(a-4)}}\mathrm{Log}\left[ \frac{\sqrt{a(a-4)}+a-2}{2} \right]\right].
\end{align}

Let us now consider the limiting cases of very small and very large ratios $a$ in Eq. (\ref{Eq:F2_Bar_0_Mass_a}). For $a \to 0$, we have
\begin{align}
     \label{Eq:F2_Bar_0_Mass_a_to_0}
      \overline{F}_2^{\rm Mass}(0,a \to 0)  &=  \frac{1}{2} -\frac{\pi}{2} a^{1/2} -\frac{\left(1+2\mathrm{Log}\left[a\right] \right)}{2}\,a + \frac{15\pi}{16} a^{3/2} +\mathcal{O}\left(a^2\right) \,,
\end{align}
which shows that, in this limit, $\overline{F}_2^{\rm Mass}(0,a)$ has a maximum value of $1/2$, as expected. In the limit $a \to \infty$, we find
\begin{align}
     \label{Eq:F2_Bar_0_Mass_a_to_Inf}
      \overline{F}_2^{\rm Mass}(0,a \to \infty)  &=  \frac{1}{3a} -\frac{12\mathrm{Log}[a]-25}{12 a^2} -
      \frac{60\mathrm{Log}[a]-97}{10 a^3}+\mathcal{O}\left(\frac{1}{a^4}\right).
\end{align}
The complete behavior of $ \overline{F}_2^{\rm Mass}(0,a)$ is shown in Fig. \ref{2D_AMM_MM_GZ_Models} (solid green line).
\subsection{Gribov--Zwanziger Model}
In the Gribov--Zwanziger (GZ) model \cite{Gribov:1977wm,zwanziger1989local} there are two contributions coming from the decomposition of the GZ gluon propagator
\begin{figure}[b]
  \centering
 \fbox{ \includegraphics[width=.45\linewidth]{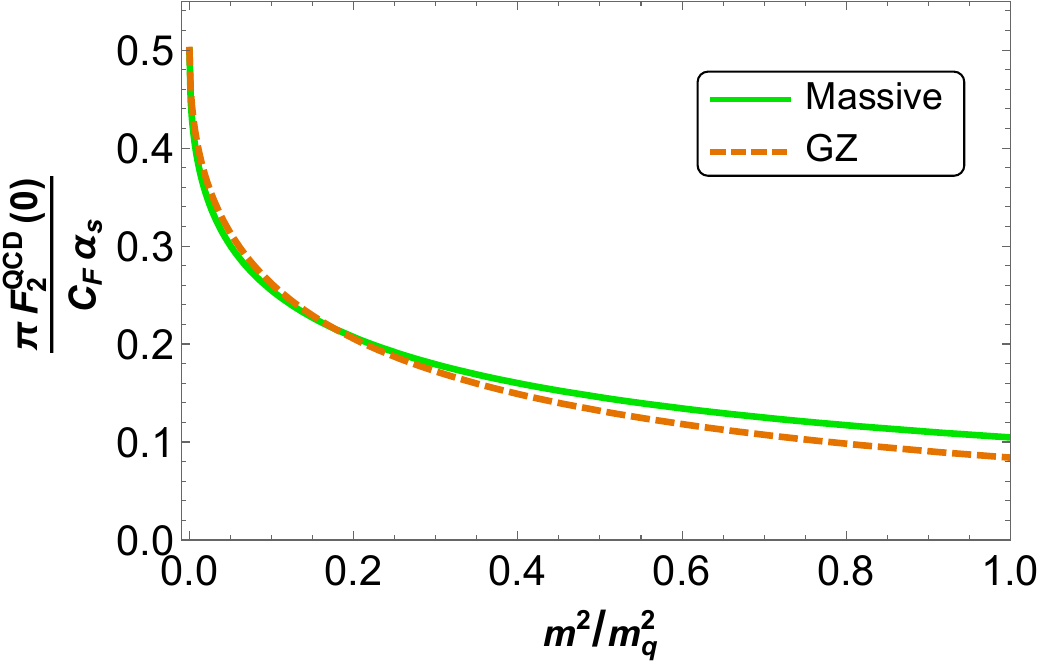}}
  \caption{$\overline{F}_2(0)$ as a function of the dimensionless ratio $m/m_q$ for the Massive model (where $m$ is the gluon mass) and the GZ model (where $m$ is the Gribov mass).}
 \label{2D_AMM_MM_GZ_Models}
\end{figure}
%
%
%
\begin{align}
\label{Eq:GZ_g_propgtr}
D_{\rm GZ}(p^2)=
\left(\frac{p^2}{p^4 + \gamma^4}\right)  =
\frac{1}{2}\left( \frac{1}{p^{2} +i\,\gamma^2} + \frac{1}{p^{2} -i\,\gamma^2} \right) .
\end{align}
\noindent
Here $\gamma$ is the Gribov mass, which is introduced in the theory to restrict the domain of the gauge functional integral to avoid multiple counting of Gribov copies. This propagator modifies the denominator in Eq. (\ref{Eq:AMM_Euc_1}), so that there will be two contributions of the form displayed in Eq. (\ref{Eu_F2_Int_Gral}), each corresponding to one of the complex-conjugated poles above, yielding $R=1/2$ and
%
%
\begin{align}
\label{Eq:Delta_mGluon_Eu_GZ}
\Delta_E\to 
\Delta_{GZ}^{\pm}\,=\, xy \,p^2 + (1-z)^2\, m_{q}^2 + z\,(\pm i\,\gamma^2)\,.
\end{align}
\noindent
At $p^2=0$, we  obtain the following one-loop form factor in the GZ model:
\begin{align}
\label{Eu_F2_p0_Int_q_V_GZ}
\overline{F_2}^{\rm GZ} (0,a)=
\frac{  \pi  F_2^{\rm GZ} (0,a) }{C_{F}\, \alpha_{s}}
\,= \frac{1}{2} \int_{0}^1 dz & \left[
\frac{ z(1-z)^2}{  (1-z)^2  + i\, z\,a} \,+   \frac{ z(1-z)^2}{  (1-z)^2  - i\, z\,a}\,\right],
\end{align}
\noindent
where $\overline{F}_2^{\rm GZ}(0,a)$ depends only on $a=(\gamma^2/m^2_{q})$, i.e. the ratio between the Gribov mass and the quark mass. Solving the integral, we obtain:
\begin{align} \nonumber
 \label{Eq:F2_Bar_0_GZ_a}
& \overline{F}_2^{\rm GZ}(0,a) =  \frac{1}{2} -\frac{a}{2} \left(  a \, \mathrm{Log}\left[ a\right] -\pi 
+ \frac{\left[(a + 2i)^2+2\right]}{\sqrt{a(a + 4i)}}  \, \mathrm{Log}\left[ \frac{ +i\,a -2 -i \sqrt{a(a + 4i)}}{2}\right]     \right. \\[5pt]
& \left.  
\hspace{172pt} - \frac{\left[(a - 2i)^2+2\right]}{\sqrt{a(a - 4i)}}  \, \mathrm{Log}\left[ \frac{ -i\,a -2 -i \sqrt{a(a - 4i)}}{2}\right]  
    \right).
\end{align}
\indent
We can also explore the limiting cases of very small and very large ratios $a$ in  $\overline{F}_2^{\rm GZ}(0,a)$. For $a \to 0$, we find
\begin{align}
     \label{Eq:F2_Bar_0_GZ_a_to_0}
      \overline{F}_2^{\rm GZ}(0,a \to 0)  &=  \frac{1}{2} -\frac{\pi}{2}\left(\frac{\sqrt{2}}{2}\right) a^{1/2} + \frac{\pi}{2}\,a - \frac{15\pi}{16} \left(\frac{\sqrt{2}}{2}\right) a^{3/2} +\mathcal{O}\left(a^2\right),
\end{align}
\noindent
showing that for  $\gamma=0$ this function has a maximum value of $1/2$, as was expected. Moreover, comparing their coefficients we can show that, for $a < 0.2$,  $\overline{F}_2^{\rm GZ}(0,a) > \overline{F}_2^{\rm Mass}(0,a)$ (see Fig. \ref{2D_AMM_MM_GZ_Models}, dashed orange line). For $a \to \infty$, we have:
\begin{align}
     \label{Eq:F2_Bar_0_GZ_a_to_Inf}
      \overline{F}_2^{GZ}(0,a \to \infty)  &=   +\frac{12\mathrm{Log}[a]-25}{12\, a^2} +
      \frac{3 \pi}{a^3}+\mathcal{O}\left(\frac{1}{a^4}\right) \,.
\end{align}
\noindent
Compared to the massive case, this expression lacks an inverse linear term, so that it decays faster, as shown in Fig. \ref{2D_AMM_MM_GZ_Models}.

From Fig. \ref{2D_AMM_MM_GZ_Models} one can see that the structure functions from the massive and GZ confining models show a similar qualitative behavior: they start from a maximum value of $1/2$ which decreases as the mass parameter of the model increases for a fixed constituent quark mass. In other words, the mass term that appears in these models has a suppressing effect on the value of $\overline{F}_2(0)$.
It may seem surprising that the GZ model, even having imaginary poles in its gluon propagator (Eq. (\ref{Eq:GZ_g_propgtr})), yields a $\overline{F}_{2}(0)$ function similar to the Massive model. This is because in its decomposition the GZ model propagator is similar to a pair of massive propagators with a conjugate counterpart which makes that the additional imaginary contributions end up canceling out leaving us with a real result and similar behavior between such functions. Even so, those additional contributions increase the value of the GZ $\overline{F}_{2}(0)$ (compared with the Massive $\overline{F}_{2}(0)$) for small values of the ``$a$'' parameter up to the value of 0.177, approximately (for $m_q=363$ MeV, $m_g=\gamma \approx 153$ MeV). For values of the a parameter greater than that, such contributions make the GZ $\overline{F}_{2}(0)$ decrease faster than the Massive $\overline{F}_{2}(0)$ (See Fig. \ref{2D_AMM_MM_GZ_Models}).

\subsection{Refined Gribov--Zwanziger Model}
The gluon propagator of the Refined Gribov-Zwanziger model can also be decomposed as a pair of massive-like propagators through the following parameterization:
\begin{eqnarray}
\label{Eq:RGZ_g_propgtr}
D_{\rm RGZ}(p^2)=
\frac{p^2 + M^2}{p^4 + (M^2+m^2)p^2 + \lambda^4 + M^2 m^2} \equiv 
%
\frac{A_{+}}{p^{2} +\alpha'_{-}} + \frac{A_{-}}{p^{2} +\alpha'_{+}} \, ,
\end{eqnarray}
\noindent
where $M^2$ and $m^2$ come from the dynamical generation of dimension 2 condensates and $\lambda^2$ is related to the Gribov mass introduced to take into account the Gribov ambiguity in the RGZ action \cite{PhysRevD.78.065047,Dudal:2010tf}. The residues $A_{\pm}$ and the poles $\alpha'_{\pm}$ can be written in terms of the RGZ mass parameters as follows:
\begin{eqnarray}
\label{fPRGZ_val}
A_{\pm}\,=\,\frac{1}{2}  \left( 1 \pm \frac{M^2 -m^2}{\sqrt{(M^{2}-m^2)^2-4\lambda^4}}\right) \equiv \frac{1}{2}\left(1\mp i \kappa\right)  \hspace{5pt},\hspace{5pt}
\\ 
\alpha'_{\pm} =\, \frac{M^2 +m^2 \pm \sqrt{(M^{2}-m^2)^2-4\lambda^4 }}{2}
\equiv \left(s\pm i t\right)
\,. \label{fPRGZ_val2}
\end{eqnarray}
\noindent
Here we defined $\kappa$ to make explicit the presence of a pair of complex-conjugated poles, since $(M^{2}-m^2)^2-4\lambda^4 < 0$, as indicated by fits to Lattice QCD data (cf. e.g. Ref. \cite{Oliveira:2012eh,DUDAL2018351}). The same fits show that $m^2 < 0$, so that it is convenient to define $\overline{m}^2 \equiv -m^2 > 0$. 
%
%

%
%
%
%
\indent
The computation of the structure function is analogous to the one for the GZ case, so that one has
\begin{align}
\label{Eq:Delta_mGluon_Eu_RGZ}
\Delta_{RGZ}^{\mp}\,=\, xy \,p^2 +  m_{q}^2\left[(1-z)^2+ z\,(a \mp ib)  \right] \,,
\end{align}
corresponding to residues $(1\mp i\kappa)/2$, where
\begin{eqnarray}
\label{FracP_RGZ_Lat_dcmpst}
\kappa\,=\, \frac{M^2 +\overline{m}^2}{\sqrt{4\lambda^4 -(M^{2}+\overline{m}^2)^2}} \hspace{5pt},\hspace{5pt}  
a =\, \frac{M^2 -\overline{m}^2 }{2m_q^2}
\hspace{5pt},\hspace{5pt}  
b =\,  \frac{ \sqrt{4\lambda^4 -(M^{2}+\overline{m}^2)^2}}{2m_q^2} \,.
\end{eqnarray}
\noindent
Following the procedure discussed for the previous models, one obtains at $p^2=0$:
\begin{align}
\label{Eu_F2_p0_Int_q_V_RGZ}
\overline{F_2}^{RGZ} (0,a,b,\kappa)
\,= \frac{1}{2} \int_{0}^1 dz  \left[
\frac{\,(1-i\kappa)\, z(1-z)^2}{  (1-z)^2 + z\,(a-ib)} \,\,+  \frac{\,(1+i\kappa)\, z(1-z)^2}{  (1-z)^2 + z\,(a+ib)} \,\right].
\end{align}
%
%
%
%
%
\noindent

The complete analytic result for the remaining integral for general values of the mass ratios $a,b$ and $\kappa$ can be found in Eq. (\ref{App:F2_Bar_0_RGZ_a}) in  Appendix \ref{App_F2_Bar_RGZ}.
Due to the presence of three independent parameters, the RGZ form factor has a more intricate structure than the Massive or GZ cases. A clearer comparison can be made if we consider extreme values for $a$ and $b$.

In the limit $a = b \to 0$, Eq. (\ref{Eu_F2_p0_Int_q_V_RGZ}) reduces to:
\begin{align}
     \label{Eq:F2_Bar_0_RGZ_a_eql_b_to_0}
& \overline{F}_2^{RGZ}(0,a  = b \to 0,\kappa)  \approx  \frac{1}{2} +\frac{\pi\sqrt{2}\left(3\kappa-1\right)}{8}\, a^{1/2} + \left(\kappa(\pi+1) + \pi-3 +2(\kappa-1)\mathrm{Log}[a]\right)\frac{a}{2}.
\end{align}
\noindent
One can see that this structure function achieves a value greater than $1/2$ for a fixed $\kappa> 1/3$ and $a\sim b\sim 0$, since its leading-order terms would be positive in contrast to the Massive and GZ cases (cf. Eqs. (\ref{Eq:F2_Bar_0_Mass_a_to_0}) and (\ref{Eq:F2_Bar_0_GZ_a_to_0})). On the other hand, for large gluon mass parameters ($a=b\to \infty$), one finds a result that is similar to the one for the Massive model, except for the additional factor of $\kappa$:
\begin{align}
     \label{Eq:F2_Bar_0_RGZ_a_eql_b_to_Inf}
    &      \overline{F}_2^{RGZ}(0,a  \sim b \to \infty,\kappa) \approx \frac{\kappa}{3a}-\frac{12\mathrm{Log}[a]-25}{12 a^2}.
\end{align}
As a consequence, the RGZ form factor decays more slowly than those of the Massive and GZ cases. Of course, setting to zero all mass parameters of the confining models, one recovers the perturbative QCD result (Eq. (\ref{Int_F2_zero_p_QCDc})).

\begin{figure}[h!]
  \centering
  \begin{tabular}[b]{c}
 \fbox{ \includegraphics[width=7.25cm, height=4.5 cm]{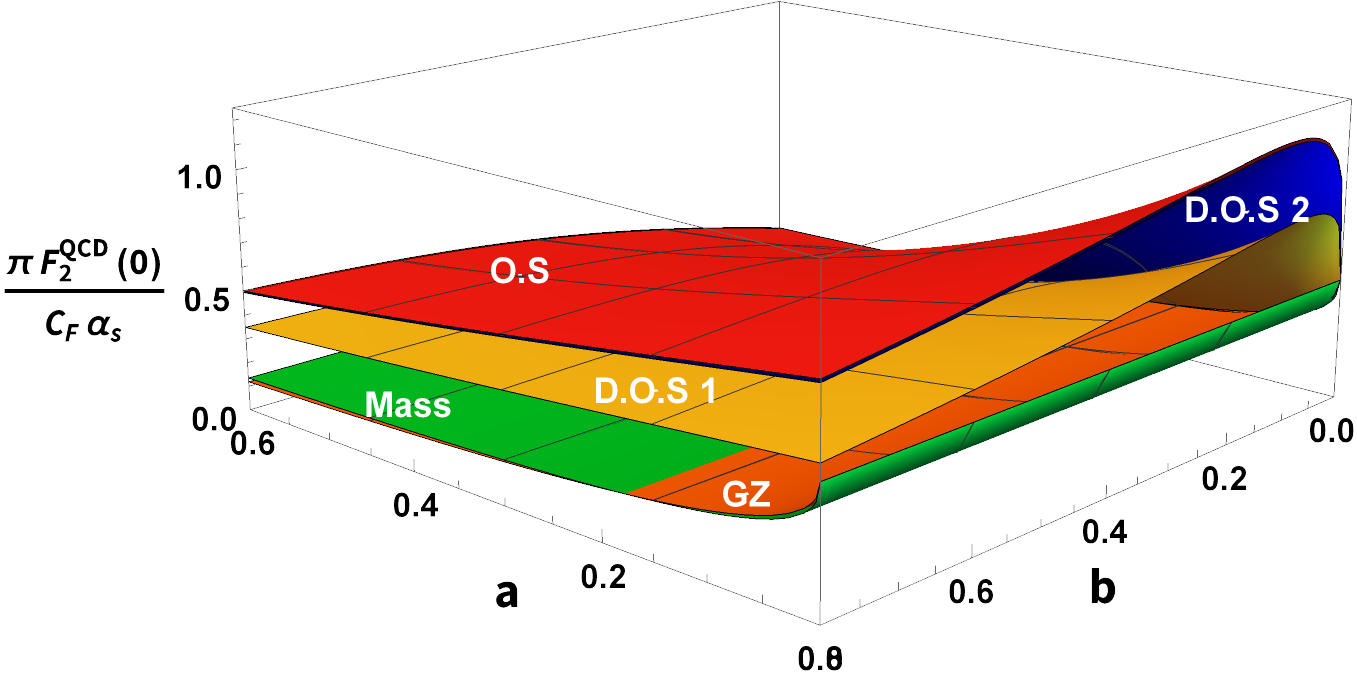}} \\
    \small (a)
  \end{tabular} \qquad
  \begin{tabular}[b]{c}
 \fbox{  \includegraphics[width=7.25cm, height=4.5 cm]{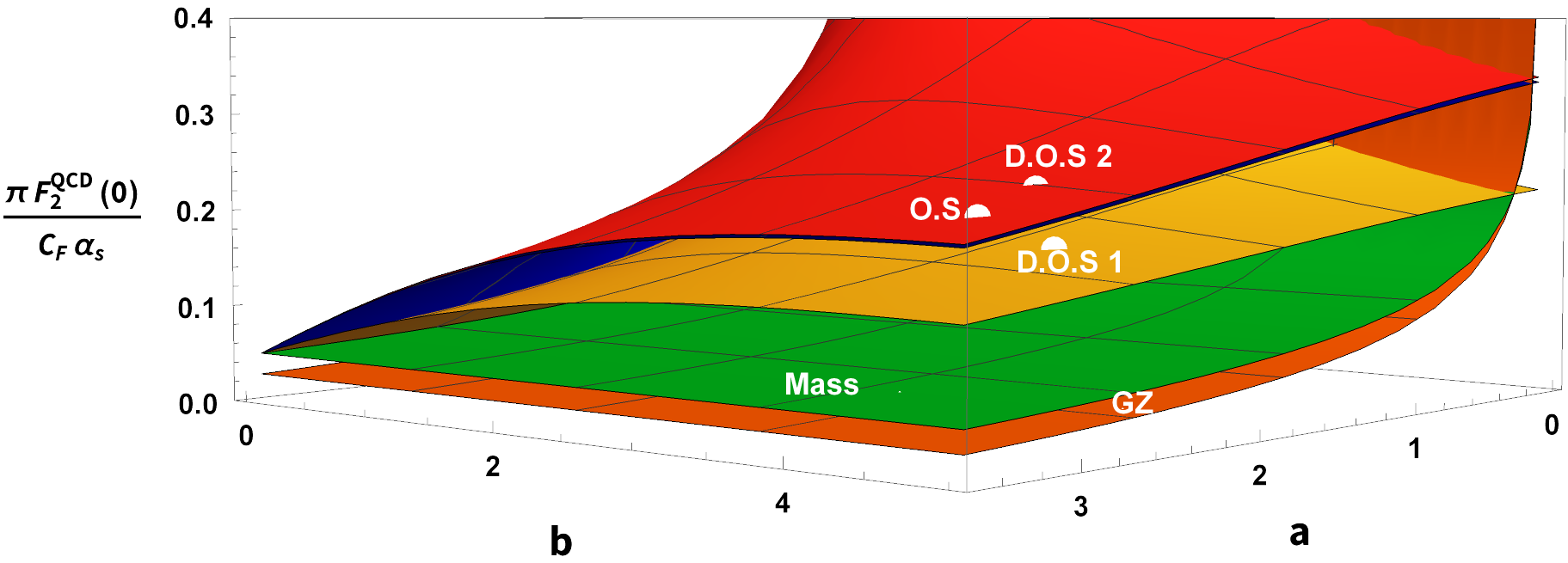}} \\
    \small (b)
  \end{tabular}
 \caption{Comparison of the behavior of $\overline{F}_2(0)$  yielded by the Massive, GZ and RGZ models with $\kappa$ fixed to lattice data (cf. Table \ref{table:1.x}). Parameters $a$ and $b$ represent ratios between confining mass parameters and the quark mass. The Massive and GZ models have no $b$ dependence.}
 \label{3D_MM_AllModels}
\end{figure}
%

%
%

\subsection{RGZ parameters from Lattice QCD}

In the RGZ model, the confining gluon propagator (Eq. (\ref{Eq:RGZ_g_propgtr})) has three independent mass parameters that give rise to complex-conjugated poles \cite{Oliveira:2012eh,DUDAL2018351}. As a consequence, the behavior of the one-loop RGZ quark form factor $\overline{F}_{2}(0)$ will be more intricate algebraically and we shall resort  to Landau-gauge Lattice gluon propagator fits to fix or at least restrict the space of parameters.
In contrast to the perturbative approach and the massive and GZ models, the RGZ $\overline{F}_2(0)$ value could be greater than 0.5 at a certain combination of small values of the $a$ and $b$ parameters (Eq. (\ref{Eq:F2_Bar_0_RGZ_a_eql_b_to_0}) and Fig. \ref{3D_MM_AllModels}.a). For large values of $a$ and $b$, the RGZ $\overline{F}_2(0)$ decreases (cf. Eq. (\ref{Eq:F2_Bar_0_RGZ_a_eql_b_to_Inf}) and Fig. \ref{3D_MM_AllModels}.b), becoming smaller than the massless gluon reference $\overline{F}_2(0)=1/2$ as the other confining models.

\begin{table}[h!]
\centering
\begin{tabular}{cc|c|c|c|c|c|c|c|}
\cline{3-9}
&&
\,\,\,  $ \mathcal{Z}_{L} $ \, \,\,&
$\,\,\,M_1^2 \,\,\,$ &
 $\,\,M_2^2\,\,$ &
  $\,\,M_3^4\,\,$ &
  \, $\,\,\,\, M\,\,\, $ \, &
\, \, $\,\,m\,\,$ \,&
\, \, $\,\,\lambda\,\,$ \, \\ \hline 
  \multicolumn{1}{ |c}{ O.S. DATA}  &&
  0.8333 & 
 4.473 & 
 0.704 & 
 0.3959 &  2.1150 & 1.9414\,i & 2.0381\\   \hline
  \multicolumn{1}{ |c}{ D.O.S. DATA 1}  && 
  1 & 
 2.525 & 
 0.510 & 
 0.2803 &  1.5890 & 1.4195\,i & 1.5222 \\ 
  \hline
  \multicolumn{1}{ |c}{ D.O.S. DATA 2}  && 0.7296 & 
 4.157 & 
 0.5922 & 
 0.3350 &  2.0389 & 1.8881\,i & 1.9730 \\\hline
\end{tabular}
\caption{RGZ parameters from Refs. \cite{Oliveira:2012eh} (O.S.) and \cite{DUDAL2018351} (D.O.S.).
}
\label{table:1.0}
\end{table}

Fits of SU(3) Lattice data for the gluon propagator provide fixed values for the RGZ parameters. For comparison, we consider the three different fits shown in Table \ref{table:1.0}. The first set is from Ref. \cite{Oliveira:2012eh}, which we call O.S. DATA, while the other two are taken from Ref. \cite{DUDAL2018351}, which we call D.O.S. 1 (with trivial gluon wave function renormalization and $\mathcal{Z}_L$) and D.O.S. 2 DATA. With this combination of values and for a constituent quark mass $m_q=363$ MeV,
the quark form factor becomes smaller than $1/2$, as displayed in Fig. \ref{3D_MM_AllModels}.b and Table \ref{table:1.x}. The similar values for O.S. and D.O.S. 2 DATA are due to the very similar values of their parameters. They share a similar (lower than one) value for the normalization factor $\mathcal{Z}_L$, which is compensated by a larger value for the parameter $M$ in the RGZ model. The D.O.S. 2 DATA has a value of $\mathcal{Z}_L$ equal to 1, which makes the value of the parameter $M$ lower than in the two previous cases. 

\begin{table}[h!]
\centering
\begin{tabular}{cc|c|c|c|c|}
\cline{3-6}
&&
$\,\,\,a\,\,\,$ &
 $\,\,b\,\,$ &
  $\,\,\kappa\,\,$ &
  $\,\,\overline{F}_2(0)\,\,$ \\ \hline
  \multicolumn{1}{ |c}{ O.S. DATA}  &&
 \,\,2.6713\,\, &
 \,\,3.9579 \,\,&
  \,\, 7.9017\,\,&
 \,\,0.2458\,\, \\   \hline
  \multicolumn{1}{ |c}{ D.O.S. DATA 1}  &&
 1.9352 &
 3.5211 &
  \,\, 4.8925 \,\,&
 0.1944 \\
  \hline
  \multicolumn{1}{ |c}{ D.O.S. DATA 2}  &&
 2.2471 &
 3.7742 &
   \,\, 7.7635 \,\,&
 0.2658  \\\hline
\end{tabular}
\caption{RGZ parameters used to calculate its $\overline{F}_2(0)$ function for a quark mass $m_q=363$ MeV.
}
\label{table:1.x}
\end{table}

\section{Nucleon magnetic moment in the Constituent Quark Model: a short review}
\label{sec:nucleonMM}

For a particle of mass $m$, spin $S$ and charge $e$ the spin magnetic moment is given by:
\begin{eqnarray}
\label{Eq:Mag_M_Def}
\boldsymbol{\mu_{part}}= \mathrm{g} \left(\frac{e}{2m}\right)\boldsymbol{S},
\end{eqnarray}
\noindent
where $\mathrm{g}$ is the Landé factor which represents the relative strength between the magnetic moment and its spin-orbit coupling.
The spin magnetic moment or just the magnetic moment of relativistic particles is derived from the fermion-photon vertex (Fig. (\ref{Fig:Q_ph_Vertex_Gral}.a)) and  represents an excellent observable to test quantum field theory corrections. Proof of that is the agreement between the 
high precision experimental measurement of the electron magnetic moment and the theoretical prediction stemming from the computation of the electron-photon vertex in perturbative QED \cite{Aoyama:2019ryr,ParticleDataGroup:2020ssz,Aoyama:2020ynm}.

%
%
%
%
%
%

It is well-known \cite{Schwartz:2014sze,Peskin:1995ev} that one can obtain the $\mathrm{g}$ factor as a function of the form factors in the fermion--photon vertex:
\begin{equation}
\label{Eq:g_factor}
\mathrm{g} = 2\left[F_1(0) + F_2(0) \right]= 2\left[1 + F_2(0) \right],
\end{equation}
\noindent
where $F_1(0)=1$ due to the electric charge conservation and $F_2$ is responsible for the anomalous contributions due to interactions, being absent at tree level.
For elementary particles at weak coupling, it is clear that the $\mathrm{g}$ factor will be very close to $2$. 

For the proton, however, the same procedure clearly does not apply. Indeed, unlike the electron ($\mathrm{g}_e \approx 2.002319$ \cite{ParticleDataGroup:2020ssz}), the experimental proton magnetic moment has a very different value ($\mathrm{g}_p \approx 5.585694$ \cite{ParticleDataGroup:2020ssz}) than the one expected for an elementary particle. This is a strong observable indication of the intricate structure of the proton, which is composed of three valence quarks and a sea of virtual quarks and gluons. 

An analytic or semi-analytic calculation of the proton magnetic moment from first principles is not available, due to the effects of the confinement phenomenon and the infrared behavior of the strong coupling. In order to obtain a prediction for observables from confining model calculations of the quark $F_2$ form factors presented in the last section, we adopt a simple yet widely used effective model that we review in what follows.

The Constituent Quark Model \cite{Perkins:1982xb,Griffiths:2008zz} describes hadron structure considering only valence quark degrees of freedom: the so-called constituent quarks, dressed by interactions and correlations with the sea partons.
As a consequence of the dressing by the interaction with the cloud of virtual quarks and gluons,  the constituent quark mass is much larger than the current mass, being about 1/3 of the mass of the nucleon for the light quarks, up ($u$) and down ($d$).

The standard CQM assumes that the quarks behave like free pointlike Dirac particles (spin $1/2$, $\mathrm{g}=2$ and electric charge $e Q_q$), so that their magnetic moment will be:
\begin{equation}
\label{Eq:MM_quarks}
\boldsymbol{\mu_{q}}= \mathrm{g} \,Q_q \left(\frac{e}{2m_q}\right)\boldsymbol{S}\, \to \mu_q=\frac{\mathrm{g}}{2}Q_q \left(\frac{M_p}{m_q}\right) \mu_N=Q_q \left(\frac{M_p}{m_q}\right) \mu_N\,,
\end{equation}
where $m_q$ is the constituent quark mass, $M_p$ is the proton mass and $\mu_N=e/2M_p$ is the nuclear magneton. For our analysis, it is important to point out that the quark magnetic moment above, Eq. (\ref{Eq:MM_quarks}), already implies the absence of quark-quark interactions, so that $F_2=0$. Naturally, this assumption will be lifted in our improved calculation of the proton magnetic moment in confining models.

A crucial parameter that is left to be determined is the constituent quark mass $m_q$. Even though it is not fixed directly from measurements, one can e.g. extract it from fittings of the hadron mass spectrum. The general procedure is to model the hadron masses within the CQM and then fit free parameters to experimental data \cite{Gasiorowicz:1981jz,Perkins:1982xb,ParticleDataGroup:2022pth}. Following Refs. \cite{Gasiorowicz:1981jz,BorkaJovanovic:2010yc} and supposing that up and down constituent quarks have the same mass $m_q$, the proton mass will be:
\begin{equation}
\label{Eq:Mproton-Hyper}
M_p= 3 m_q +\frac{a'}{m_q^2} \, 4 \sum_{i<j}^{3}{\bf{S}}_i\cdot{\bf S}_j= 3 m_q -3 \frac{a'}{m_q^2} 
\,,
\end{equation}
where the second term corresponds to the binding energy, with ${\bf S}_i$ being the spin operator of the $i$-th quark in the proton and $a'$ is a free parameter. The specific expression of this contribution is obtained as a hyperfine splitting coming from a Coulomb-type  color potencial interaction between pairs of constituent quarks. A combined fit of the baryon octet mass spectrum yields $\frac{a'}{m_q^2} \approx 50 \, {\rm MeV}$, so that, using the proton mass $M_p = 938.2720$ MeV \cite{ParticleDataGroup:2020ssz} in Eq.(\ref{Eq:Mproton-Hyper}),
we obtain the constituent quark mass $m_q=363$ MeV.


%
%
Finally, in the CQM, the proton magnetic moment comes from vector sums of the magnetic moments of its constituent quarks. Using the totally symmetric SU(6) (flavor SU(3) $\times$ spin SU(2)) wavefunctions for the three-quark states, one predicts for the proton:
\begin{equation}
\label{Eq:CQM_proton_MM}
\mu_{p}=  \frac{4}{3} \mu_{u} - \frac{1}{3}\mu_{d}\,.
\end{equation}
\noindent
In what follows, we use this setup to investigate how confining gluon propagators affect this observable.

%
%
%
%

%
%
\section{
Corrections to the proton magnetic moment from confining models
}
\label{sec:confining-MM}
%
%
%
%

\indent

From Eq. (\ref{Eq:g_factor}) and the results of Section \ref{sec:results-confining} for the $F_2$ form factor in different  models, the quarks g factors can be obtained directly, including effects from confinement in the anomalous contributions:
\begin{eqnarray}
\mu_{q} =\,Q_q \left(\frac{e}{2m_q}\right)\frac{1}{2}\left[2\left(1+F_2^{q}(0)\right]\right)=Q_q \left(\frac{M_p}{m_q}\right)\left[1+F_2^{q}(0)\right] \mu_N \,,
\end{eqnarray}
\noindent
where $Q_u=+2/3$, $Q_d=-1/3$, $m_u=m_d=m_q=363$ MeV, $M_p=938.2720$ MeV, and $\mu_N=e/2M_p$. The various models of QCD interactions will enter through the result for the $F_2^q$ form factors of the quark-photon vertex. 

For one-loop perturbative QED and QCD (Eqs. (\ref{Int_F2_zero_QEDc}) and (\ref{Int_F2_zero_p_QCDc})), the magnetic moment of the quark will be ($C_F=(N^2-1)/(2N)=4/3$):
\begin{eqnarray}
\mu_{q} =  \,Q_q \left(\frac{M_p}{m_q}\right)\left(1+Q_q^{2} \left(\frac{\alpha}{2\pi}\right) + C_{F}\left(\frac{\alpha_{s}}{2\pi}\right) \right) \, \mu_N.
\end{eqnarray}

For the confining models used in our analysis (Eqs. (\ref{Int_F2_zero_QEDc}), (\ref{F2_Int_pzero_V_qmgq}), (\ref{Eu_F2_p0_Int_q_V_GZ}) and (\ref{Eu_F2_p0_Int_q_V_RGZ})), we have:
%
%
%
\begin{eqnarray}
\label{Eq:quark_amm_Conf_Mod}
\mu_{q} =  \,Q_q \left(\frac{M_p}{m_q}\right)\left(1+Q_q^{2} \left(\frac{\alpha}{2\pi}\right) + C_{F}\left(\frac{\alpha_{s}}{\pi}\right)\overline{F}_2 (0)  \right) \, \mu_N\,,
\end{eqnarray}
where $\overline{F}_2 (0)$ depends on the ratio of the confining models' masses vs the constituent quark mass. 

It is now straightforward to write, using Eq.(\ref{Eq:CQM_proton_MM}), the proton magnetic moment for the standard constituent quark model:
\begin{equation}
\mu_{p}^{CQM} =  \frac{4}{3} \mu_{u} - \frac{1}{3}\mu_{d}=\left[ \frac{4}{3} \,Q_u\left(\frac{M_p}{m_q}\right) - \frac{1}{3}\,Q_d \left(\frac{M_p}{m_q}\right) \right]\mu_N= \frac{M_p}{m_q}\, \mu_N\,,
\end{equation}
as well as an improved expression, including anomalous contributions from QED and QCD interactions:
\begin{eqnarray}
\label{Eq:proton_amm_Conf_Mod}
\mu_{p}^{+QFT} =\mu_{p}^{CQM}  \left[  1+\left(\frac{4}{3}Q^3_u -\frac{1}{3}Q^3_d
\right)\frac{\alpha}{2\pi} + \left[\,C_F\,\overline{F}_2 (0) \,\right]\frac{\alpha_s}{\pi}
\right]\,.
\end{eqnarray}
%
The QED anomalous term will be fixed to the well-known value ($\alpha=1/137$), and we concentrate on how the QCD contribution $\propto \overline{F}_2(0)$ affects the equation above within different approaches. 

\subsection{\label{sec:alphaS} Running Coupling in the deep IR}

To obtain a quantitative prediction for the proton magnetic moment a crucial parameter that we have not yet discussed is the strong coupling constant $\alpha_s$ appearing in Eq.(\ref{Eq:proton_amm_Conf_Mod}). A consistent calculation implies that one has actually a running coupling evaluated at the typical scale of the process: $p^2\to 0$ in this case. 

The deep IR limit of the strong running coupling is however unknown and very hard to define in the nonperturbative region. Even in standard perturbative QCD at high energies, the running coupling is a scheme-dependent quantity that relies heavily on observable input. The most up-to-date determination of $\alpha_s$ uses information e.g. from $\tau$ decay data described with next-to next-to next-to leading order (${\rm N}^3{\rm LO}$) predictions \cite{ParticleDataGroup:2022pth}, yielding a value of $\alpha_s(p\approx 2 \,{\rm GeV})\approx 0.3$. It is thus reasonable from the perturbative renormalization group evolution to expect a  value larger than $\alpha_s=0.3$ as one decreases the momentum below $2$ GeV. Nevertheless, in the deep IR one has to resort to nonperturbative approaches and different models provide scenarios varying from a vanishing $\alpha_s$ to a diverging one as the scale goes to zero. A large collection of results can be found in Ref. \cite{Deur:2016tte}, where it becomes clear that a variety of descriptions arrive at a saturating running coupling, with $0\lesssim\alpha_s(0)\lesssim 4$. 

 For the upper values in this range, $\alpha_s \approx 4$, a diagrammatic one-loop technique could only be justified if the effective expansion parameter is actually lower than $\alpha_s$, as suggested by the convergence and success of perturbation theory within the Curci-Ferrari model for correlation functions of gluons and ghosts in $SU(3)$ \cite{Tissier:2011ey,Gracey:2019xom}. There, the effective expansion parameter is argued to be $\lambda_{\rm CF}=3\alpha_s/4\pi $. Even though a detailed analysis of this type is still not available for GZ and RGZ theories, the good qualitative results at tree level and a few one-loop calculations encourage the same assumption.

In what follows, we fix $\alpha_s$ within the interval $[0,4]$ and compare different confining models for the same coupling.

\subsection{\label{sec:level_CQM_Mm} Results for the massive model}

\begin{figure}[b]
  \centering
 
    \begin{tabular}[b]{c}
 \fbox{ \includegraphics[width=.44\linewidth]{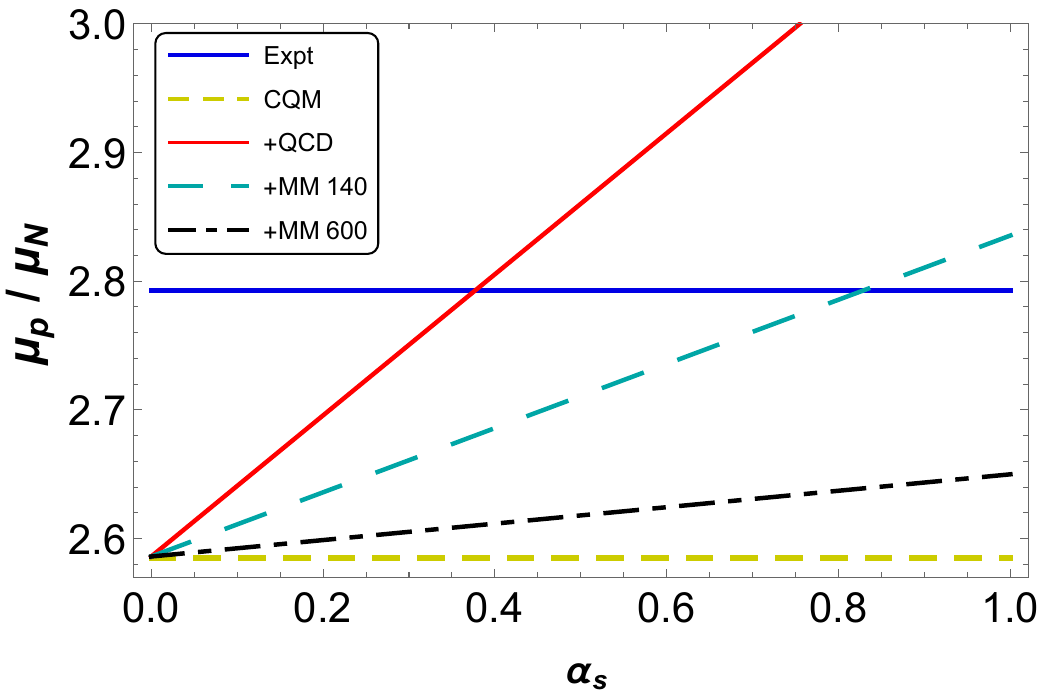}} \\
    \small (a)
  \end{tabular} \qquad
  \begin{tabular}[b]{c}
 \fbox{  \includegraphics[width=.44\linewidth]{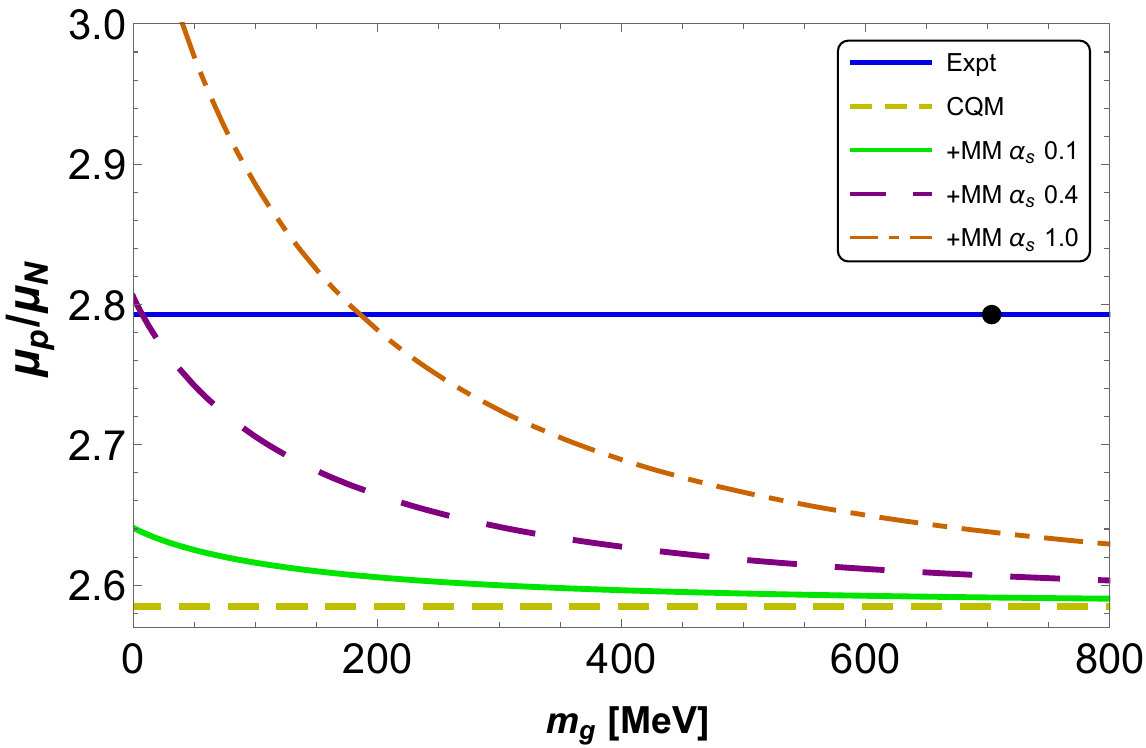}} \\
    \small (b)
  \end{tabular}
\caption{Comparison between the experimental (Expt) proton magnetic moment $\mu_p$, the CQM result and that from CQM modified by MM. The zero mass limit is labeled ``$+$QCD''. (a) $\mu_p$ vs $\alpha_s$. (b) $\mu_p$ vs $m_g$. The black point corresponds to $\alpha_s=4$ and $m_g=703.83$ MeV. 
} \label{MM_QCD_Mass_mass_alph_s}
\end{figure}

Let us now analyze the behavior of the proton magnetic moment in the CQM modified with corrections from the Massive gluon model (MM) at one-loop order for the quark form factor $F_2$. The final expression is given by Eq.(\ref{Eq:proton_amm_Conf_Mod}) with $\overline{F}_2(0)$ given by
Eq. (\ref{Eq:F2_Bar_0_Mass_a}), where $a=(m_g/m_q)^2$, with $m_q=363$ MeV.

Fig.  \ref{MM_QCD_Mass_mass_alph_s}
displays the MM results for the proton magnetic moment in units of the nucleon magneton as a function of the strong coupling constant $\alpha_s$ and the gluon mass $m_g$. One can notice
that, for any fixed nonzero coupling, the MM provides a larger proton magnetic moment with respect to the free/standard CQM prediction, becoming closer to the measurement for low $\alpha_s$. As can be seen in the analytic limit shown in Eq.(\ref{Eq:F2_Bar_0_Mass_a_to_Inf}), the  absence of interactions, i.e. the standard CQM, is recovered when the gluon mass becomes extremely large.
Indeed, in Fig. \ref{MM_QCD_Mass_mass_alph_s}.b, the gluon mass is shown to decrease the value of $\mu_p$.
Moreover, low values of the strong coupling, $\alpha_s<0.38$, render the MM corrections too small to attain the experimental magnetic moment in this setup. Nevertheless, the overall discrepancy is less than $10\%$ for this small $\alpha_s$ regime which could still be considered reasonable for an effective description.

On the other hand, for strong couplings, $\alpha_s>0.38$, there is always a nonzero gluon mass that reproduces the experimental value.
We collect a few parameter sets ($m_g,\alpha_s$)
that provide a MM prediction for the proton magnetic moment in agreement with data in the first lines of Tables \ref{table:2.0} and \ref{table:3.0}.


%
%
%
%
%
\begin{table}[t]
\centering
\begin{tabular}{cc|c|c|c|c|}
\cline{3-6}
&& \,\, $ m$ [MeV]\,\, &
$\,\,\,\,\,0 \,\,\,\,\,$ &
 $\,\,140\,\,$ &
  $\,\,600\,\,$  \, \\ \hline 
  \multicolumn{1}{ |c}{ Mass}  &&
   \multirow{2}{*}{$\alpha_s\,\, \| \,\, \lambda_{CF} $} & 
  \multirow{2}{*}{$\,\,0.38\,\, \| \,\, 0.091\,\,$} & 
$\,\,0.83\,\, \| \,\, 0.198\,\,$& 
$\,\,3.24\,\, \| \,\, 0.773\,\,$ \\ \cline{1-2} \cline{5-6}
  \multicolumn{1}{ |c}{ GZ}  && 
   & 
   & 
$\,\,0.82\,\, \| \,\, 0.196\,\,$& 
$\,\,5.30\,\, \| \,\, 1.265\,\,$ \\\hline
\end{tabular}
\caption{Parameter set where results from QCD, the Massive model and the GZ model agree with the experimental proton magnetic moment $\mu_p$. The mass $m$ stands for the gluon mass $m_g$ in the MM and the Gribov mass $\gamma$ in GZ. 
}
\label{table:2.0}
\end{table}

\begin{table}[h!]
\centering
\begin{tabular}{cc|c|c|c|c|c|}
\cline{3-7}
&&
\,\,\,  $ \alpha_{s} \,\, \| \,\, \lambda_{CF}$ \, \,\,&
$\,\,\,0.38 \,\, \| \,\, 0.091\,\,\,$ &
 $\,\,0.40  \,\, \| \,\, 0.095\,\,$ &
  $\, 1.00 \,\, \| \,\, 0.239\,\,$ &
  \, $\,\,\,\,4.00 \,\, \| \,\, 0.955\,\,\, $  \, \\ \hline 
  \multicolumn{1}{ |c}{ Mass}  &&
  $m_g$ [MeV] & 
 0 & 
 7.30 & 
 185.64 & 703.83\\   \hline
  \multicolumn{1}{ |c}{ G.Z.}  && 
  $\gamma $ [MeV]& 
 0 & 
 9.70 & 
 179.77 &  516.80 \\\hline
\end{tabular}
\caption{Parameter set where results from QCD, the Massive model and the GZ model agree with the experimental value of $\mu_p$. $\alpha_s$ was fixed. $\lambda_{CF}=3\alpha_s/4\pi$ is the expansion parameter of the Curci--Ferrari (massive) model.}
\label{table:3.0}
\end{table}


%
%
%
%
%
\subsection{\label{sec:level_CQM_GZm} Results for the Gribov--Zwanziger model}

\begin{figure}[ht]
  \centering
    \begin{tabular}[b]{c}
 \fbox{ \includegraphics[width=.44\linewidth]{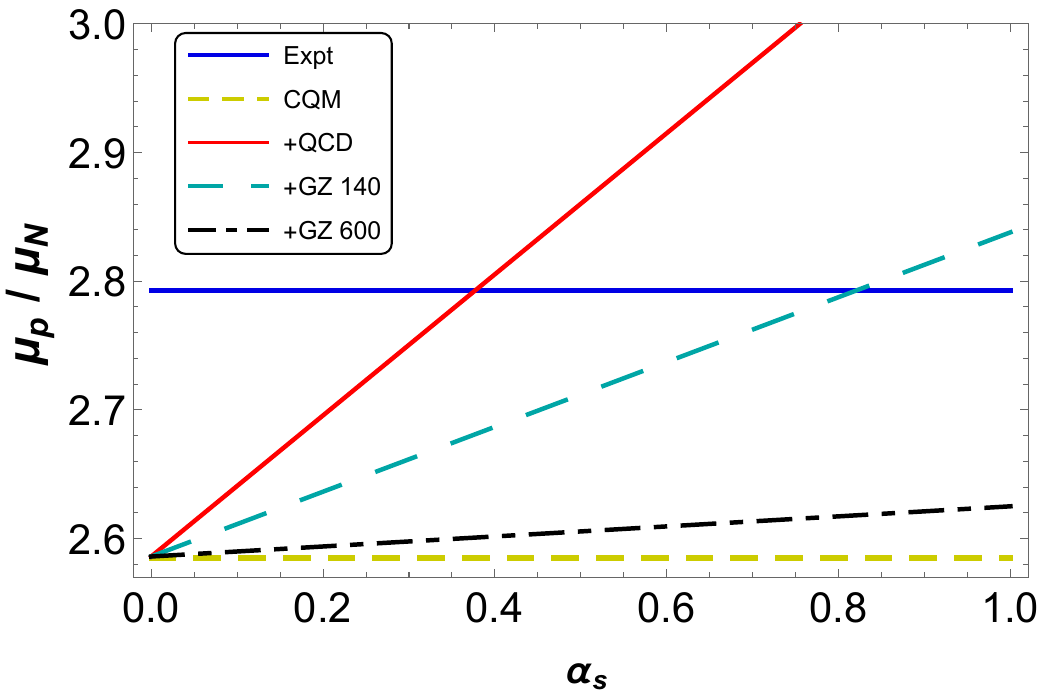}} \\
    \small (a)
  \end{tabular} \qquad
  \begin{tabular}[b]{c}
 \fbox{  \includegraphics[width=.44\linewidth]{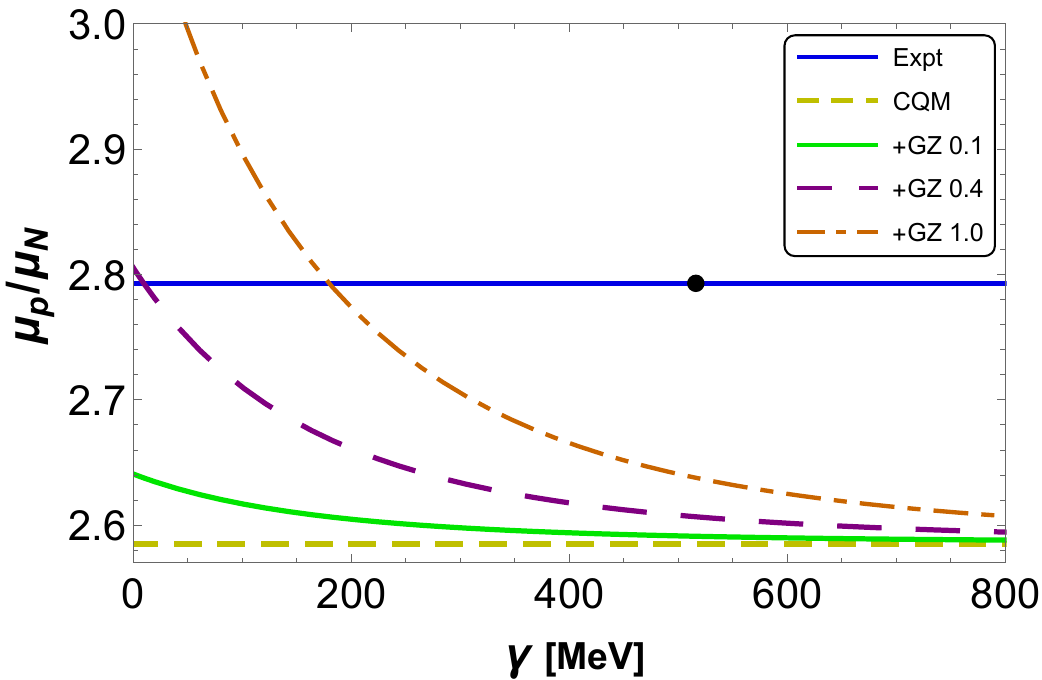}} \\
    \small (b)
  \end{tabular}
  \caption{Comparison between the experimental (Expt) proton magnetic moment ($\mu_p$), the CQM result and that from CQM modified by GZ. (a) $\mu_p$ vs $\alpha_s$ (b) $\mu_p$ vs $m_g$. The black point corresponds to $\alpha_s=4$ and $\gamma=516.80$ MeV.}
 \label{MM_QCD_GZ_mass_alph_s}
\end{figure}

Turning to the Gribov-Zwanziger confining model and using again the Eq. (\ref{Eq:proton_amm_Conf_Mod}), where $\overline{F}_2(0)$ now comes from Eq. (\ref{Eu_F2_p0_Int_q_V_GZ}), we analyze the GZ prediction for the proton magnetic moment.
Since the GZ $\overline{F}_2(0)$ behaves very similarly to the $\overline{F}_2(0)$ of the massive model (cf. Fig. \ref{2D_AMM_MM_GZ_Models}), it is straightforward to conclude that the magnetic moment coming from the GZ model behaves in a similar way to the massive model in this fixed coupling setup. In fact, that can be verified by comparing Figs. \ref{MM_QCD_Mass_mass_alph_s} and \ref{MM_QCD_GZ_mass_alph_s}. Moreover, it is important to notice that the presence of a set of imaginary poles in the gluon propagator brought about by Gribov quantization does not bring any nonphysical behavior to the proton magnetic moment obtained in this approximation.

Fig. \ref{MM_QCD_GZ_mass_alph_s}.a displays the proton magnetic moment in units of the nucleon magneton as a function of the strong coupling $\alpha_s$. It shows a larger value for the calculated magnetic moment with a massless gluon (solid, oblique red line) and GZ corrections than for the $\mu_p^{CQM}$. As in the massive case, the larger the confining mass scale introduced in the gluon propagator, the smaller is the modification with respect to the free CQM model, which is recovered when the Gribov mass $\gamma\to \infty$.

In Tables \ref{table:2.0} and \ref{table:3.0}, we can appreciate the quantitative difference between the massive and GZ parameters that reproduce the measured proton magnetic moment. In Fig. \ref{2D_AMM_MM_GZ_Models}, the GZ form factor $\overline{F}_2(0)$ is shown to be slightly smaller (larger) than the massive result for masses lower (higher) than $\approx 153$ MeV. Since the proton magnetic moment depends on the combination $\overline{F}_2(0)\,\alpha_s$, the GZ model requires stronger couplings for Gribov masses larger than $141$ MeV, while the massive model remains in the $\lambda_{CF}<1$ region even for gluon masses around $600$ MeV.

\subsection{\label{sec:level_CQM_RGZm} Results for the Refined Gribov--Zwanziger model}

\begin{figure}[b]
  \centering
  \begin{tabular}[b]{c}
 \fbox{  \includegraphics[width=.44\linewidth]{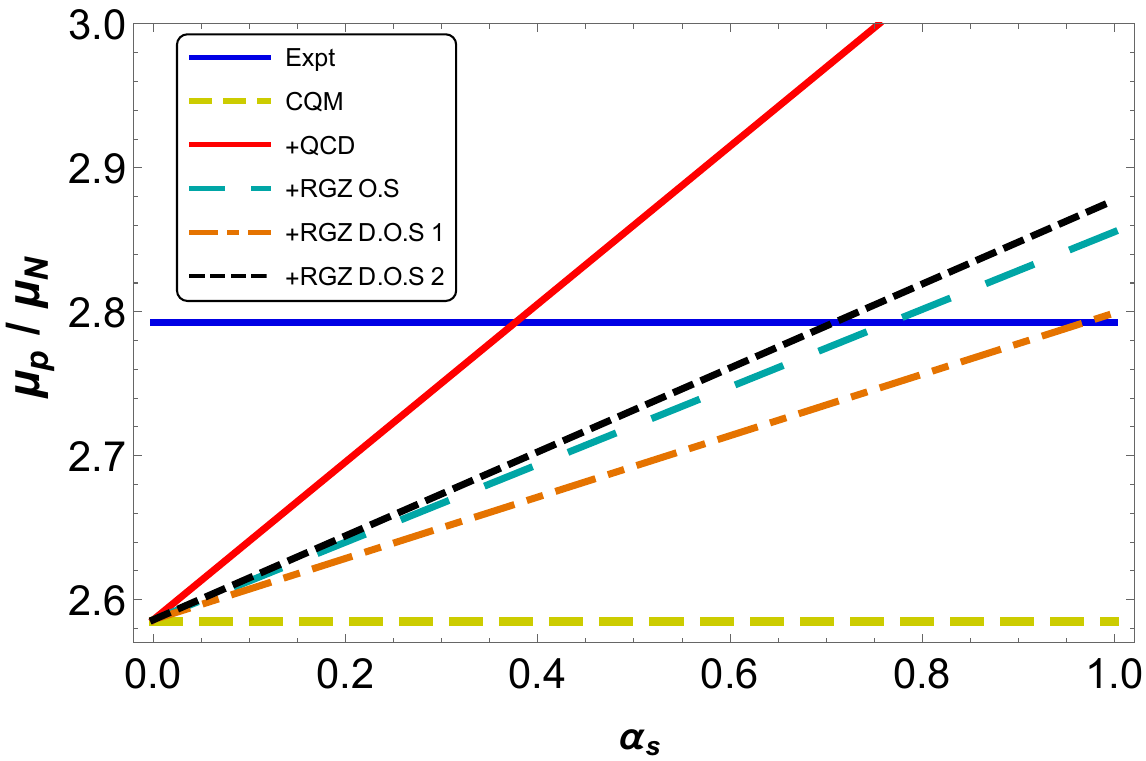}} 
  \end{tabular}
  \caption{Comparison between the experimental (Expt) proton magnetic moment ($\mu_p$), the CQM result and that from CQM modified by RGZ contributions as a function of the coupling constant $\alpha_s$.}
 \label{MM_QCD_RGZ_alph_s}
\end{figure}

Finally, for the Refined--Gribov--Zwanziger model, the one-loop result for the form factor $\overline{F}_2(0)$ is given in Eq. (\ref{Eu_F2_p0_Int_q_V_RGZ})
and the proton magnetic moment in the RGZ-improved CQM framework is computed as before using the general expression in Eq. (\ref{Eq:proton_amm_Conf_Mod}). As discussed previously, the RGZ model has three parameters that can be adjusted to fit the lattice gluon propagator. For the moment we will use the lattice QCD values that we call O.S., D.O.S. 1 and D.O.S. 2 DATA (cf. Tables \ref{table:1.0} and \ref{table:1.x}). The corresponding results for the proton magnetic moment are shown in Fig. \ref{MM_QCD_RGZ_alph_s}.

\begin{figure}[b]
  \centering
  \begin{tabular}[b]{c}
 \fbox{ \includegraphics[width=.44\linewidth]{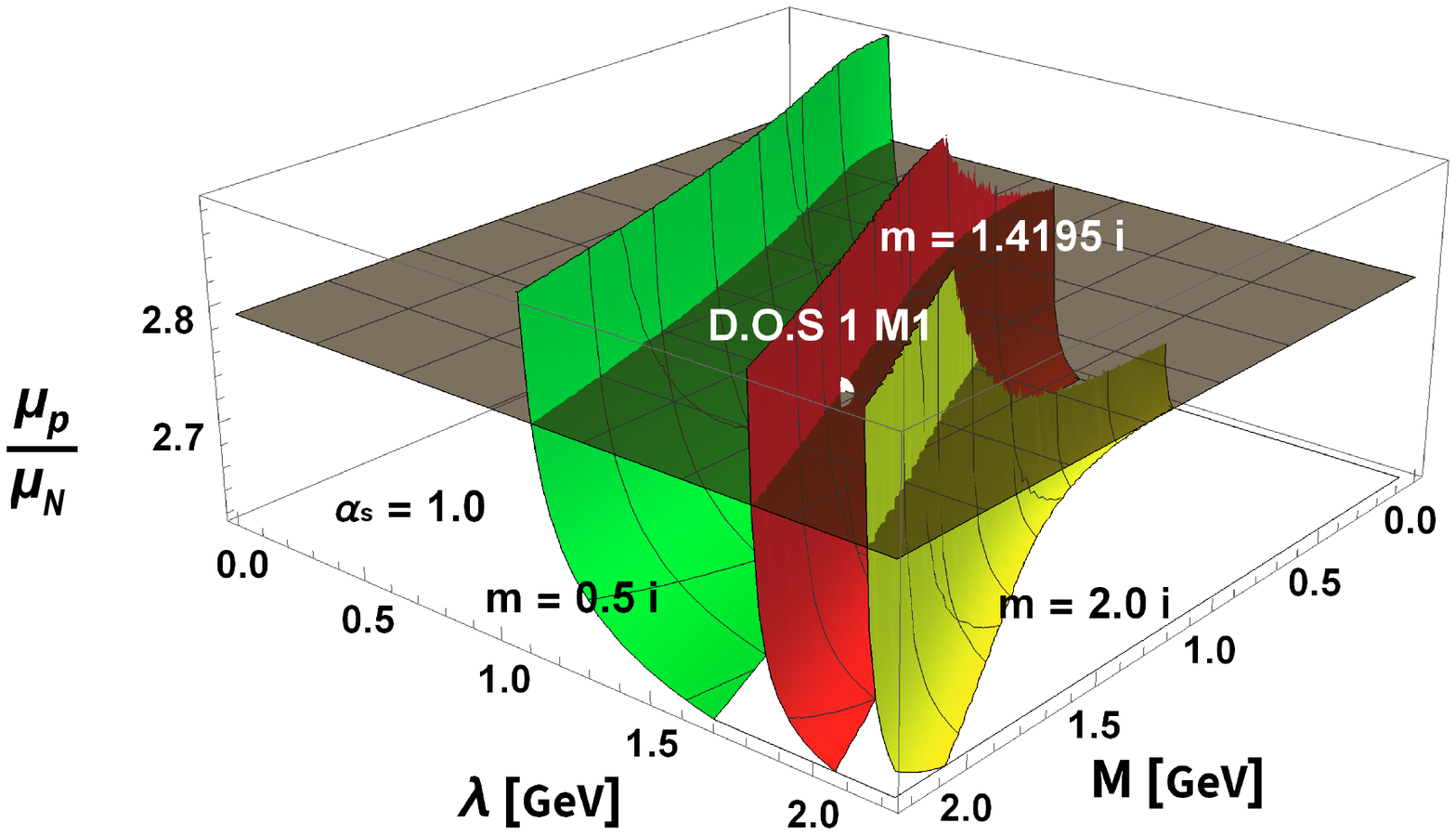}} \\
    \small (a)
  \end{tabular} \qquad
  \begin{tabular}[b]{c}
 \fbox{  \includegraphics[width=.44\linewidth]{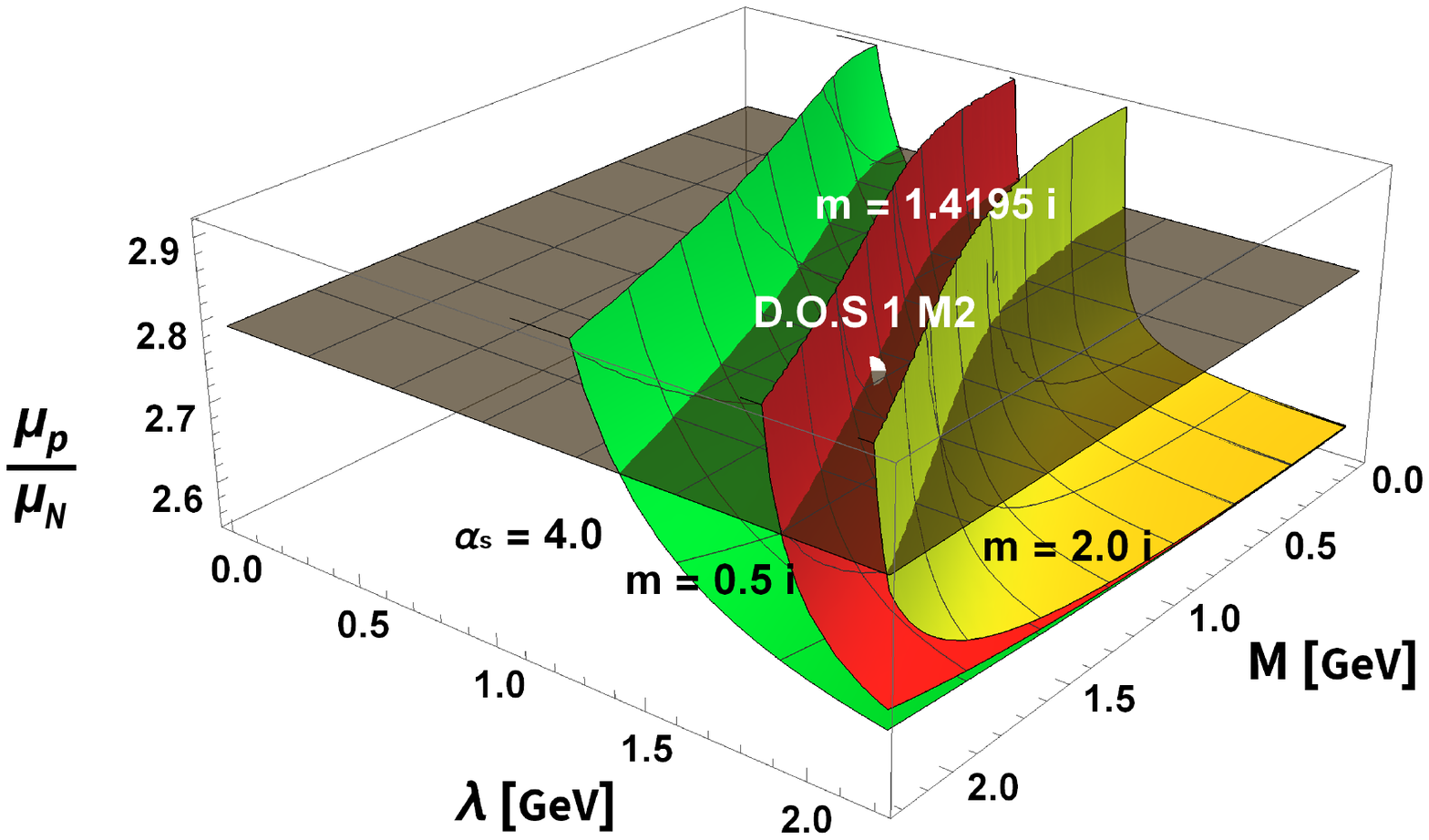}} \\
    \small (b)
  \end{tabular}
    \begin{tabular}[b]{c}
 \fbox{ \includegraphics[width=.43\linewidth]{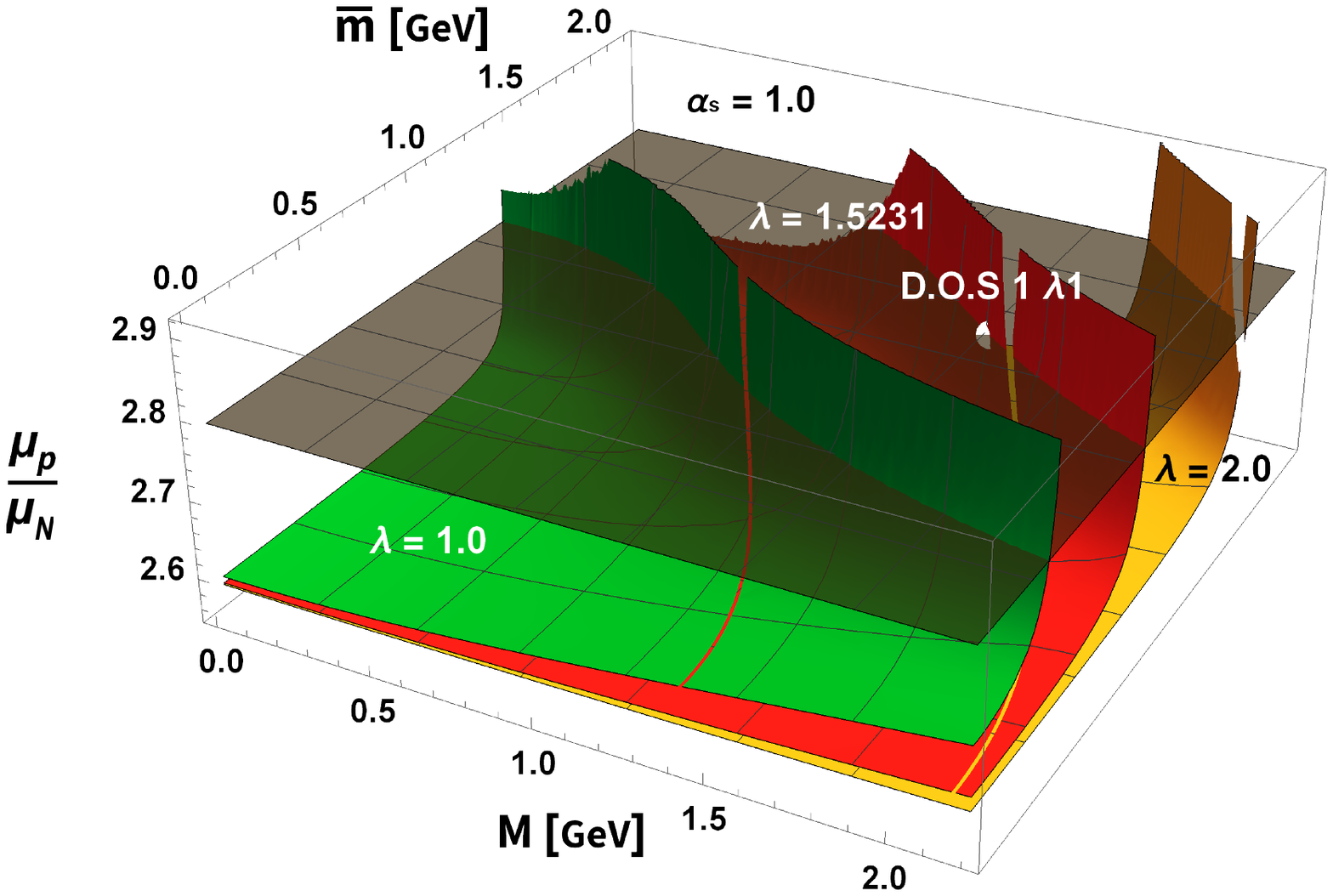}} \\
    \small (c)
  \end{tabular} \qquad
  \begin{tabular}[b]{c}
 \fbox{  \includegraphics[width=.43\linewidth]{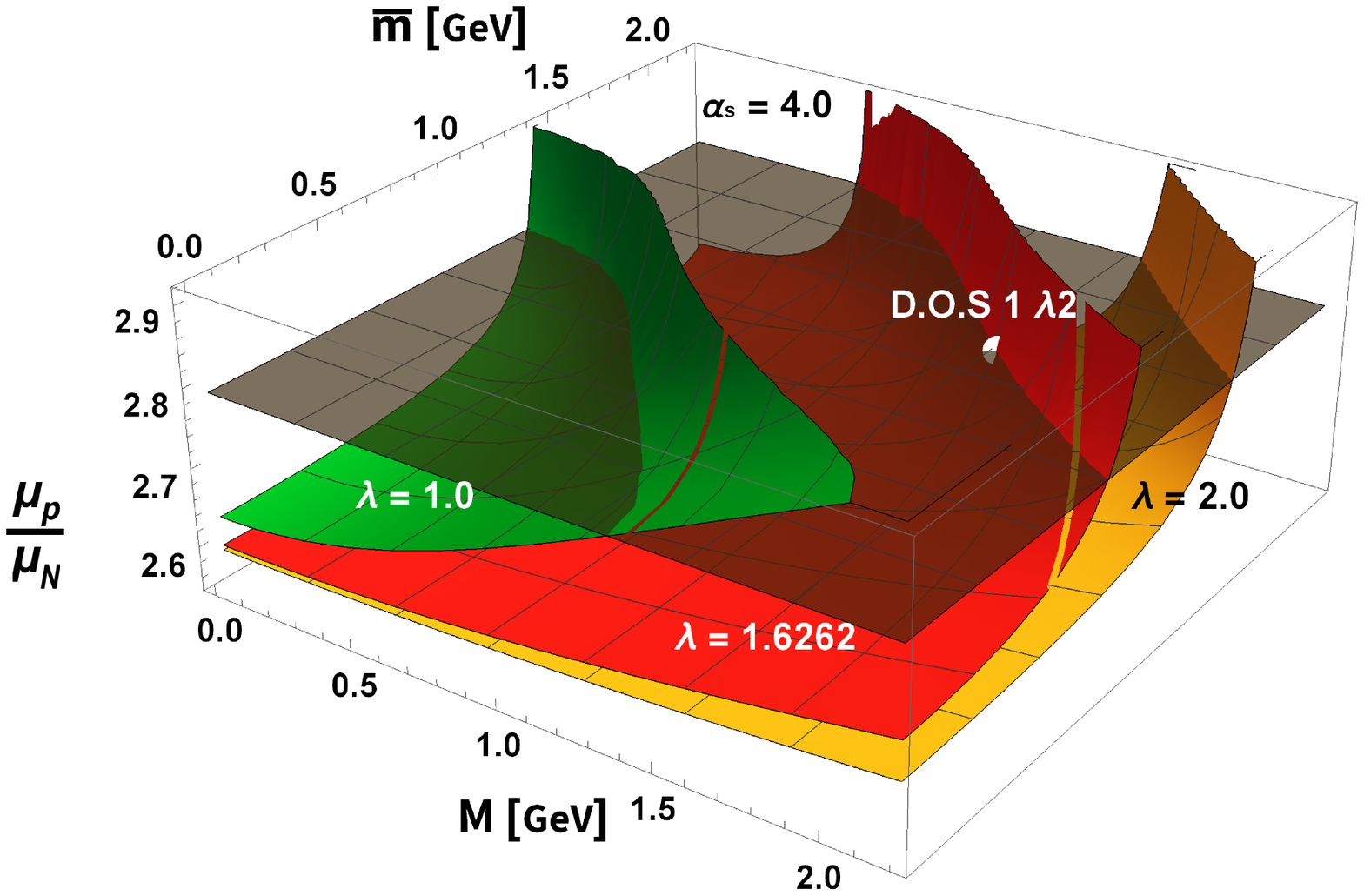}} \\
    \small (d)
  \end{tabular}
  \caption{$\mu_p$ as a function of different RGZ parameters ($M,m,\lambda$). For $m$ fixed: (a) $\alpha_s=1$ and (b) $\alpha_s=4$. For $\lambda$ fixed: (c) $\alpha_s=1$ and (d) $\alpha_s=4$, where $\overline{m}=- i\, m$. 
The black plane corresponds to the experimental proton magnetic moment $\mu_p^{Expt}$. 
The white dots correspond to the case in which two mass parameters assume the values of the Lattice fits, while the third one can be found in (Table \ref{table:4.x}).}
 \label{MM_3D_RGZ_DOS_MOD}
\end{figure}

The general features brought about by confining mass scales in the gluon propagator are also seen in the RGZ results, namely: (i) corrections from interactions increase the proton magnetic moment with respect to the free CQM prediction, (ii) the modification is smaller than the massless gluon limit, and (iii) there are strong coupling values that render the RGZ prediction equal to the experimental value.
In particular, it is interesting to note that, within this setup, fixing the three mass parameters from Landau-gauge Lattice QCD data we are able to describe quantitatively the Lattice gluon propagator, the proton mass and the proton magnetic moment with $\alpha_s=0.85 \pm 0.15$.


We can also explore further the parameter space of the RGZ model, relaxing the requirement that the corresponding gluon propagator represents the best fit of available Lattice data.
The behavior of the magnetic moment in the CQM-RGZ for different values of $\{M,m,\lambda\}$ is displayed in Fig. \ref{MM_3D_RGZ_DOS_MOD}. It is clear that there many mass parameter sets that reproduce the measured proton magnetic moment, so that this observable alone, i.e. without the Lattice gluon propagator information, does not impose a strong constraint on the RGZ parameters.

\begin{table}[t]
\centering
\begin{tabular}{|cc|c|c|c|}
 \hline
%
  \multicolumn{1}{|c}{\,\,  $ \alpha_{s} \,\, \| \,\, \lambda_{CF}$ \, \,}   &&
$\,\,0.40  \,\, \| \,\, 0.095\,\,$ &
$\, 1.00 \,\, \| \,\, 0.239\,\,$ &
$\,\,\,\,4.00 \,\, \| \,\, 0.955\,\,\, $\\ \hline 
\multicolumn{1}{ |c}{$\{m,\lambda\}$ \,\,\,fixed $\|\,\, M$ }  &&
 \,\,1.6234\,\, & 
 \,\,1.5870\,\,& 
  \,\,1.3662\,\,\\   \hline
\multicolumn{1}{ |c}{$\{M,\lambda\}$ \,\,fixed $\|\,\, m$}  && 
 1.4487\,$i$ & 
 1.4178\,$i$ & 
 1.2013\,$i$  \\   \hline
\multicolumn{1}{ |c}{$\{M,m\}$ fixed $\|\,\, \lambda$}&& 
 1.5064 & 
 1.5231 & 
 1.6262 \\\hline
\end{tabular}
\caption{RGZ parameters used to calculate the $\mu_p$ for a quark mass $m_q=363$ MeV. Fixed $\{M,m,\lambda\}$ values come from D.O.S. 1 (Ref. \cite{DUDAL2018351}).
}
\label{table:4.x}
\end{table}
%
%
%




In Fig. \ref{MM_3D_RGZ_DOS_MOD}, we have chosen strong coupling values in the range discussed in subsection \ref{sec:alphaS} to illustrate the results for larger couplings as well. We have $\alpha_s=1.0$ in Figs. \ref{MM_3D_RGZ_DOS_MOD}.a and \ref{MM_3D_RGZ_DOS_MOD}.c and $\alpha_s=4.0$ in 
Figs. \ref{MM_3D_RGZ_DOS_MOD}.b and \ref{MM_3D_RGZ_DOS_MOD}.d.
Furthermore, in Table \ref{table:4.x}
we show the quantitative difference in the parameters when one of the mass parameters in the propagator is allowed to change to reproduce the measured proton magnetic moment for different values of $\alpha_s$. Overall, the  modification in the parameters is no greater than $10\%$, which is still a reasonable ballpark for the approximations made in the current description.



\section{SUMMARY AND outlook}
\label{sec:summary}

Confining models based on infrared modifications of the gluon propagator by the presence of mass scales have been used to describe different correlation functions of quarks and gluons with reasonable success when compared to results from Lattice simulations. In this paper we have focused on the quark-photon vertex and the QCD anomalous contribution to the quark magnetic moment to obtain predictions within these confining models for the observable proton magnetic moment.

For that matter, we have computed the one-loop contributions to the quark-photon vertex using gluon propagators associated with the massive or Curci-Ferrari-like model, the Gribov-Zwanziger model and the Refined Gribov-Zwanziger model. To construct the proton magnetic moment from the quark form factor, we have adopted the simple, yet widely used Constituent Quark Model as a general framework for comparison between the interactions mediated by different confining gluon descriptions.

The first important observation that can be extracted from our results is that the confinement ingredients here introduced did not bring qualitative changes to the observable predicted. This statement should not be taken for granted, especially since GZ and RGZ present  a nonstandard analytical structure in the form of complex-conjugated gluon poles.
Moreover, all three models display the general features of increasing the predicted value for the proton magnetic moment with respect to the free CQM case. This enhancement caused by the presence of confining mass scales in the gluon correlation function is however smaller than a na\"ive calculation with a massless gluon propagator would provide. Furthermore, quantitative differences between the various confining models can be sizable, so that observable constraints would in principle be possible.

We recall that the setup used throughout the paper involves mainly the following parameters: (i) the constituent quark mass, (ii) the strong coupling constant in the deep IR limit, and (iii) confining mass scales (one, in the MM and GZ cases, and three in the RGZ model). Therefore, in order to have constraints from the observed proton magnetic moment on confining mass parameters, one needs sufficient information on the constituent quark mass and the strong coupling constant. The former is better known and we have fixed from the observed baryon octet mass spectrum (including the proton mass), using a CQM description in the presence of hyperfine interactions stemming from Coulomb-like potentials between pairs of quarks inside the proton. The latter, however, is much less under control, since a nonperturbative unique definition of the QCD running coupling is not available. We have therefore shown results for a set of finite $\alpha_s(0)$ values inspired by a variety of descriptions, collected in Ref. \cite{Deur:2016tte}. On the other hand, if the QCD running coupling goes to zero in the deep IR, as suggested e.g. in the gluon-sector of the renormalization-group-improved Curci-Ferrari model in the infrared-safe scheme, the anomalous QCD contribution to the quark -- and thus the proton -- magnetic moments are of course trivial.
It is important to note, however, that the relevant coupling here is the one for the quark sector, which is probably larger than the one in the gluon (and ghost) interactions \cite{Pelaez:2017bhh}. Finally, the proton magnetic moment could be a good observable to obtain information on the behavior of the running coupling constant in the deep IR, so that the further development of predictions from infrared QCD models is called for.

There are several directions of improvement to be investigated in the future. Our results can be straightforwardly extended to full momentum dependence and other form factors of the quark-photon vertex to be confronted with Dyson-Schwinger equations \cite{Chang:2010hb} and future Lattice results \cite{Leutnant:2018dry}.
Higher loop calculations in the setup presented here within each of the confining models would be useful to verify the convergence of perturbation theory in these nontrivial backgrounds for an actual observable instead of gauge-dependent correlation functions of quarks and gluons. 

\acknowledgments
%
This work was partially supported by CAPES (Finance Code 001), Conselho Nacional de Desenvolvimento Cient\'{\i}fico e Tecnol\'{o}gico (CNPq), Funda\c c\~ao Carlos Chagas Filho de Amparo \` a Pesquisa do Estado do Rio de Janeiro (FAPERJ), and INCT-FNA (Process No. 464898/2014-5).

%
%

\begin{appendix}

\section{Refined Gribov-Zwanziger $\overline{F}_2$(0) integral  \label{App_F2_Bar_RGZ}}
Unlike the Massive and GZ case, the RGZ function will be characterized by two parameters, $a$, and $b$, which will be a ratio between the RGZ mass terms and the quark mass as was detailed in the Eqs. (\ref{FracP_RGZ_Lat_dcmpst}) and (\ref{Eu_F2_p0_Int_q_V_RGZ}):
\begin{align}
    \overline{F}_2^{RGZ}(0,a,b,\kappa) = \frac{1}{2}\,\int_{0}^1 dz  \left[
\frac{\,(1-i\kappa)\, z(1-z)^2}{  (1-z)^2 + z\,(a-ib)} \,\,+  \frac{\,(1+i\kappa)\, z(1-z)^2}{  (1-z)^2 + z\,(a+ib)} \,\right],
\end{align}
%
%
%
%
\noindent
where as in the previous cases we separate the equation into integrable parts and integrate it for an arbitrary z what will gives us an expression like:
\begin{align}
\label{App:F2_Bar_Int_z_RGZ}
  \overline{F}_2^{RGZ}(0,a,b,\kappa) \to \frac{z^2}{2} + z\,\left(b\,\kappa -a\right) + \frac{(1\mp i\kappa)}{2}\, \left( \left(a \mp ib \right) \frac{z\left( a \mp ib -2 \right) +1  }{  (1-z)^2 + z\,(a \mp ib)} \right)
\end{align}
\noindent
which was rewritten in a convenient form to obtain the solution of the RGZ $\overline{F}_2(0,a)$ function:
\begin{align}\nonumber
 \label{App:F2_Bar_0_RGZ_a}
  & \overline{F}_2^{RGZ}(0,a,b,\kappa) = \frac{1}{2} + \left(b\,\kappa -a\right) + \\[5pt] \nonumber
&  \frac{(1- i\kappa)}{2}\,
 \left(  \frac{\left( a - ib\right) \left(a - ib -2 \right) }{2}  \right) 
 \left( \frac{\mathrm{Log}\left[ a^2 +b^2 \right] }{2} -i\,\mathrm{Tan}^{-1}\left[\frac{b}{a}\right] \right)  + \\[5pt] \nonumber
&  \frac{(1+ i\kappa)}{2}\, \left(  \frac{\left( a + ib\right) \left(a + ib -2 \right) }{2}  \right) 
 \left( \frac{\mathrm{Log}\left[ a^2 +b^2 \right] }{2} +i\,\mathrm{Tan}^{-1}\left[\frac{b}{a}\right] \right) - \, \\[5pt]\nonumber
&   
\frac{(1- i\kappa)}{2} \left(   \frac{\left( a - ib\right) \left[(a - ib -2)^2 -2 \right] }{2 \sqrt{(a - ib -2)^2-4}}  \right) \left(
%
 \mathrm{Log} \left[\frac{\sqrt{(a - ib -2)^2-4} +a-ib-2}{2} \right]  \right)- \\[5pt] 
& \frac{(1+ i\kappa)}{2} \left(   \frac{\left( a + ib\right) \left[(a + ib -2)^2 -2 \right] }{2 \sqrt{(a + ib -2)^2-4}}  \right) \left(
%
 \mathrm{Log} \left[\frac{\sqrt{(a + ib -2)^2-4}+a+ib-2}{2} \right]   \right),
\end{align}
\noindent
where for RGZ data from QCD lattice (see Table \ref{table:1.x}):
\begin{eqnarray}
\kappa\,=&&\, \frac{M^2 +\overline{m}^2}{\sqrt{4\lambda^4 -(M^{2}+\overline{m}^2)^2}} \geq 0 \hspace{5pt},\hspace{5pt}  
 a\,=\, \frac{s}{m_q^2} \geq 0 \hspace{5pt},\hspace{5pt} 
  b\,=\, \frac{t}{m_q^2}  \geq 0
\end{eqnarray}
\begin{align}
s =\, \frac{M^2 -\overline{m}^2 }{2} \geq 0 \hspace{10pt}\& \hspace{10pt} t =\,  \frac{ \sqrt{4\lambda^4 -(M^{2}+\overline{m}^2)^2}}{2} \geq 0 .
\end{align}

We will also explore the borderline cases of RGZ $\overline{F}_2(0,a)$ (Eq. (\ref{App:F2_Bar_0_RGZ_a})) for $a \to 0$ and  $a \to \infty$, such that:
\begin{align}\nonumber
    & \overline{F}_2^{RGZ}(0,a \to 0,b \to 0,\kappa)  =  \frac{1}{2} +\frac{\pi}{2}\left(\frac{\sqrt{2}}{2}\right)(\kappa-1)\, b^{1/2} + \frac{b}{2}\left(\pi +\kappa(1+2\mathrm{Log}[b])\right) +\mathcal{O}\left(b^{3/2}\right) \\[5pt] \nonumber
    & \hspace{35pt} a\left( -\frac{\pi}{4}\left(\frac{\sqrt{2}}{2}\right)  \frac{\kappa+1}{b^{1/2}} +\frac{\pi \kappa-3-2\mathrm{Log}[b]}{2}  -\frac{45\pi}{32}\left(\frac{\sqrt{2}}{2}\right)(\kappa-1) \,b^{1/2} \right. \\[5pt] 
    & \hspace{125pt}\left. + \frac{8\kappa-3\pi-6\kappa\mathrm{Log}[b]}{6}\,b  +\mathcal{O}\left(b^{3/2}\right) \right) +\mathcal{O}\left(a^2 \right),
\end{align}
\noindent
whereas at the other end we have that:
\begin{align}\nonumber
    &      \overline{F}_2^{RGZ}(0,a \to \infty,b \to \infty,\kappa) = \frac{1}{3a} +\frac{1}{a^2} \left( \frac{b\,\kappa}{3} -\frac{12\mathrm{Log}[b]-25}{12} \right)\\[5pt]
     & \hspace{35pt}    +\frac{1}{a^3}   \left( -\frac{b^2}{3} -\frac{180\mathrm{Log}[a]-291}{30} -\frac{b\,\kappa}{30}\left(60\mathrm{Log}[a]-155\right)\right) +\mathcal{O}\left( \frac{b^3}{a^4}\right).
\end{align}
\end{appendix}
%

\bibliographystyle{apsrev4-1} 
\bibliography{refs}

\end{document}